\documentclass[journal, 10pt, twocolumns]{IEEEtran}

\usepackage{cite}
\usepackage{graphicx}%
\usepackage{multirow}%
\usepackage{amsmath,amssymb,amsfonts}%
\usepackage{amsthm}%
\usepackage{mathrsfs}%
\usepackage{xcolor}%
\usepackage{textcomp}%
\usepackage{manyfoot}%
\usepackage{booktabs}%
\usepackage{algorithm}%
\usepackage{algorithmicx}%
\usepackage{algpseudocode}%
\usepackage [latin1]{inputenc}
\usepackage{listings}%
\usepackage{textcomp}
\usepackage{xcolor}
\usepackage{mathtools}
\usepackage{makecell}
\usepackage[font={small}]{caption}
\usepackage{caption}
\usepackage{subcaption}
\usepackage{cleveref}
\usepackage{optidef}
\usepackage{breqn}
\usepackage{mathrsfs}
\usepackage{accents}
\usepackage{acronym}
\usepackage{soul}
\usepackage{bm}
\usepackage{url}
\usepackage{wrapfig}
\usepackage{todonotes}
\usepackage{verbatim}
\usepackage{pdfpages}
\usepackage{siunitx}
\usepackage{tikz}
\usepackage{cancel}
\usepackage{pgfplots}
\pgfplotsset{compat=1.3}
\usepgfplotslibrary{statistics, groupplots}
\usepackage{tikz}
\usepackage{pgfplotstable}
\usetikzlibrary{shapes, arrows, positioning, calc, backgrounds, angles, quotes, patterns}
\usetikzlibrary{external}
\graphicspath{{./figures/}}
\acrodef{prop}[\textit{MIMORPH}]{MIMO Radio Platform for Heterogeneous wireless systems}

\acrodef{3gpp}[3GPP]{3rd Generation Partnership Project}
\acrodef{6g}[6G]{Sixth Generation}
\acrodef{abft}[A-BFT]{Association Beamforming Training}
\acrodef{ack}[ACK]{Acknowledge}
\acrodef{adc}[ADC]{Analog-to-Digital Converter}
\acrodef{aoa}[AoA]{Angle of Arrival}
\acrodef{aod}[AoD]{Angle of Departure}
\acrodef{ap}[AP]{Access Point}
\acrodef{amc}[AMC]{Advanced Mezzanine Card}
\acrodef{awv}[AWV]{Antenna Wave Vector}
\acrodef{axi}[AXI]{Advanced eXtensible Interface}
\acrodef{ber}[BER]{Bit Error Rate}
\acrodef{bft}[BFT]{Beamforming Training}
\acrodef{bp}[BP]{Backprojection}
\acrodef{brp}[BRP]{Beam Refinement Phase}
\acrodef{ca}[CA]{Carrier Aggregation}
\acrodef{cs}[CS]{Compressed Sensing}
\acrodef{cdf}[CDF]{Cumulative Distribution Function}
\acrodef{cef}[CEF]{Channel Estimation Field}
\acrodef{cfo}[CFO]{Carrier Frequency Offset}
\acrodef{cir}[CIR]{Channel Impulse Response}
\acrodef{cfr}[CFR]{Channel Frequency Response}
\acrodef{csi}[CSI]{Channel State Information}
\acrodef{csirs}[CSI-RS]{CSI-Reference Signal}
\acrodef{cs}[CS]{Compressed Sensing}
\acrodef{cv}[CV]{Constant Velocity}
\acrodef{cnn}[CNN]{Convolutional Neural Network}
\acrodef{nmpm}[NMPM]{Normalized Multiband Peak Magnitude}
\acrodef{empw}[EMPW]{Empirical Multiband Peak Width}
\acrodef{msps}[MSPS]{Mega-Samples per Second}
\acrodef{dft}[DFT]{Discrete Fourier Transform}
\acrodef{dl}[DL]{Deep Learning}
\acrodef{dma}[DMA]{Direct Memory Access}
\acrodef{dmg}[DMG]{Directional Multi Gigabit}
\acrodef{dti}[DTI]{Data Transfer Interval}
\acrodef{edmg}[EDMG]{Enhanced Directional Multi Gigabit}
\acrodef{ekf}[EKF]{Extended Kalman Filter}
\acrodef{elu}[ELU]{Exponential-Linear Unit}
\acrodef{fmcw}[FMCW]{Frequency-Modulated Continuous-Wave}
\acrodef{fov}[FOV]{Field-of-View}
\acrodef{ft}[FT]{Fourier Transform}
\acrodef{fr1}[FR1]{Frequency Range 1}
\acrodef{fr2}[FR2]{Frequency Range 2}
\acrodef{fr3}[FR3]{Frequency Range 3}
\acrodef{gpio}[GPIO]{General Purpose Input/Output}
\acrodef{gsps}[GSPS]{Giga-Samples per Second}
\acrodef{gtd}[GTD]{Geometrical Theory of Diffraction}
\acrodef{har}[HAR]{Human Activity Recognition}
\acrodef{ht}[HT]{High Throughput}
\acrodef{idft}[IDFT]{Inverse Discrete Fourier Transform}
\acrodef{if}[IF]{Intermediate Frequency}
\acrodef{ifs}[IFS]{Inter-Frame Spacing}
\acrodef{iht}[IHT]{Iterative Hard Thresholding}
\acrodef{ista}[ISTA]{Iterative Shrinkage-Thresholding Algorithm}
\acrodef{isac}[ISAC]{Integrated Sensing And Communication}
\acrodef{isafs}[ISAFS]{Iterative Spatial Ambiguity Function Subtraction}
\acrodef{jcs}[JCS]{Joint Communication and Sensing}
\acrodef{jpdaf}[JPDAF]{Joint Probabilistic Data Association Filter}
\acrodef{los}[LoS]{Line-of-Sight}
\acrodef{lo}[LO]{Local Oscillator}
\acrodef{lbm}[LBM]{Loop-Back Memory}
\acrodef{qam}[QAM]{Quadrature Amplitude Modulation}
\acrodef{mae}[MAE]{Mean Absolute Error}
\acrodef{mcs}[MCS]{Modulation and Coding Scheme}
\acrodef{md}[$\mu$D]{micro-Doppler}
\acrodef{mimo}[MIMO]{Multiple Input Multiple Output}
\acrodef{mmwave}[mmWave]{Millimeter-Wave}
\acrodef{msps}[MSPS]{Mega-Samples per Second}
\acrodef{mu}[MU]{Multiple User}
\acrodef{MUSIC}[MUSIC]{MUlti SIgnal Classification}
\acrodef{mpc}[MPC]{Multiband Phase Coherence}
\acrodef{nac}[NAC]{Normalized Auto Correlation}
\acrodef{nco}[NCO]{Numerical Controlled Oscillator}
\acrodef{nlos}[NLoS]{Non-Line-of-Sight}
\acrodef{nn}[NN]{Neural Network}
\acrodef{nls}[NLS]{Nonlinear Least-Squares}
\acrodef{ofdm}[OFDM]{Orthogonal Frequency Division Multiplexing}
\acrodef{omp}[OMP]{Orthogonal Matching Pursuit}
\acrodef{ospa}[OSPA]{Optimal Subpattern Assignment}
\acrodef{ota}[OTA]{Over-the-Air}
\acrodef{per}[PER]{Packet Error Rate}
\acrodef{phy}[PHY]{Physical Layer}
\acrodef{pl}[PL]{Programmable Logic}
\acrodef{pov}[POV]{Point-of-View}
\acrodef{ps}[PS]{Processing System}
\acrodef{po}[PO]{Phase Offset}
\acrodef{pri}[PRI]{Pulse Repetition Interval}
\acrodef{pslr}[PSLR]{Peak-to-Sidelobe Ratio}
\acrodef{spbp}[SPBP]{Subsets Product Backprojection}
\acrodef{ransac}[RANSAC]{Random Sample Consensus}
\acrodef{raf}[RAF]{Range Ambiguity Function}
\acrodef{rmse}[RMSE]{Root Mean Squared Error}
\acrodef{rf}[RF]{Radio Frequency}
\acrodef{rfsoc}[RFSoC]{Radio Frequency System on a Chip}
\acrodef{rcs}[RCS]{Radar Cross-Section}
\acrodef{rss}[RSS]{Received Signal Strength}
\acrodef{rom}[ROM]{Read Only Memories}
\acrodef{rx}[RX]{receiver}
\acrodef{smi}[SMI]{Standard Multistatic Imaging}
\acrodef{sc}[SC]{Single Carrier}
\acrodef{sdr}[SDR]{Software Defined Radio}
\acrodef{siso}[SISO]{Single Input Single Output}
\acrodef{sls}[SLS]{Sector Level Sweep}
\acrodef{snr}[SNR]{Signal-to-Noise Ratio}
\acrodef{soc}[SoC]{System on a Chip}
\acrodef{spb}[SPB]{Signal Processing Blocks}
\acrodef{srrc}[SRRC]{Square-Root-Raised-Cosine}
\acrodef{ssb}[SSB]{Synchronization Signal Block}
\acrodef{ssr}[SSR]{Super Sample Rate}
\acrodef{sta}[STA]{Station}
\acrodef{std}[STD]{Standard Deviation}
\acrodef{stf}[STF]{Short Training Field}
\acrodef{stft}[STFT]{Short Time Fourier Transform}
\acrodef{su}[SU]{Single User}
\acrodef{sar}[SAR]{Synthetic Aperture Radar}
\acrodef{tf}[TF]{Time-Frequency}
\acrodef{to}[TO]{Timing Offset}
\acrodef{toa}[ToA]{Time of Arrival}
\acrodef{tof}[ToF]{Time of Flight}
\acrodef{tx}[TX]{transmitter}
\acrodef{ue}[UE]{User Equipment}
\acrodef{ula}[ULA]{Uniform Linear Array}
\acrodef{usrp}[USRP]{Universal Software Radio Peripheral}
\acrodef{vht}[VHT]{Very High Throughput}
\acrodef{vna}[VNA]{Vector Network Analyzer}
\acrodef{uwb}[UWB]{Ultra-Wide Band}
\acrodef{wlan}[WLAN]{Wireless Local Area Network}
\acrodef{wrc}[WRC]{World Radio Congress}

\newcommand{\eq}[1]{Eq.~\eqref{#1}}

\newcommand{\fig}[1]{Fig.~\ref{#1}}

\newcommand{\secref}[1]{Section~\ref{#1}}

\newcommand{\mytexttilde}{{\raise.17ex\hbox{$\scriptstyle\mathtt{\sim}$}}}

\IEEEaftertitletext{\vspace{-2.5\baselineskip}}

\begin{document}

\title{Toward Multiband Sensing in FR3: Frequency Anisotropy Characterization and Non-Contiguous Bands Aggregation Algorithms}

\author{Jacopo Pegoraro, Gianmaria Ventura, Dario Tagliaferri, Marco Mezzavilla, Andrea Bedin, Michele Rossi, Joerg Widmer
\thanks{Jacopo Pegoraro, Gianmaria Ventura, and Michele Rossi are with the Department of Information Engineering, University of Padova, Italy (\mbox{e-mail}: jacopo.pegoraro@unipd.it). Dario Tagliaferri and Marco Mezzavilla are with the Department of Electronics, Information, and Bioengineering, Politecnico di Milano, Italy. Andrea Bedin and Joerg Widmer are with the IMDEA Networks Institute, Spain.\\
This work has received funding from the Smart Networks and Services Joint Undertaking (SNS JU) under the European Union's Horizon Europe research and innovation programme, project MultiX (Grant Agreement No 101192521).}
}

\maketitle

\begin{abstract}
\ac{fr3} in the 7-24 GHz band will be the new spectrum for \acs{6g} wireless networks. The bandwidth availability and diversity of \ac{fr3} offer unprecedented opportunities for coherent multiband \ac{isac}, which aggregates the carrier phase information from multiple frequency bands to increase the sensing resolution to the cm-level.
However, the frequency anisotropy of sensing targets over GHz-wide bands and the non-contiguity of the \acs{6g} spectrum, pose critical challenges to the application of existing multiband \ac{isac} techniques.
We present the first study on coherent multiband sensing in \ac{fr3}. We experimentally characterize the frequency anisotropy of targets and propose new phase coherence metrics for multiband processing. 
Then, we analyze the impact of non-contiguous \ac{fr3} bands considered by \acs{3gpp}, and design a new algorithm to mitigate the resulting sensing artifacts, outperforming existing techniques.
Our results represent a first step toward fully developing multiband \ac{isac} for \ac{fr3}.
\end{abstract}


\section{Introduction}\label{sec:intro}

There is a general consensus that \acf{isac} will be one of the key pillars of \ac{6g} cellular networks~\cite{prelcic2025six}, endowing them with radar-like capabilities to sense the location and movements of objects and people in the environment.
However, the limited spectrum availability and the non-contiguity of the frequency bands allocation represent a bottleneck that prevents the sensing accuracy and resolution of \ac{isac} systems from achieving their cm-level theoretical value. 
These limitations have recently sparked the research on \textit{coherent multiband} \ac{isac}~\cite{Wan_OFDM_multiband_ISAC_VTM, li2025multi}, which leverages the \textit{carrier phase} information contained in the channel state information acquired in multiple, possibly non-contiguous, frequency bands to increase the delay and range resolution. This line of work naturally aligns with the opportunities offered by the \acf{fr3} spectrum, spanning approximately $7$-$24$~GHz.

The \ac{fr3} spectrum has emerged as a `goldilocks' range for \ac{6g}, striking a balance between the favorable propagation characteristics of FR1 and the abundant bandwidth of FR2 mmWave bands~\cite{bazzi2025upper}. Unlike sub-$6$~GHz, \ac{fr3} offers much larger contiguous and non-contiguous allocations, which are critical to support the surge in mobile data traffic and artificial intelligence-driven applications anticipated by $2030$. At the same time, its propagation loss and penetration characteristics remain far more manageable than those in mmWave, enabling coverage in urban and suburban areas with existing tower infrastructure. This unique combination makes \ac{fr3} a key enabler for high-capacity, mid-range coverage with resilience to blockage~\cite{kang2024fr3cellular}.

\begin{figure}[!t]
    \centering
    \includegraphics[trim={3cm 4cm 2cm 4cm},clip,width=\columnwidth]{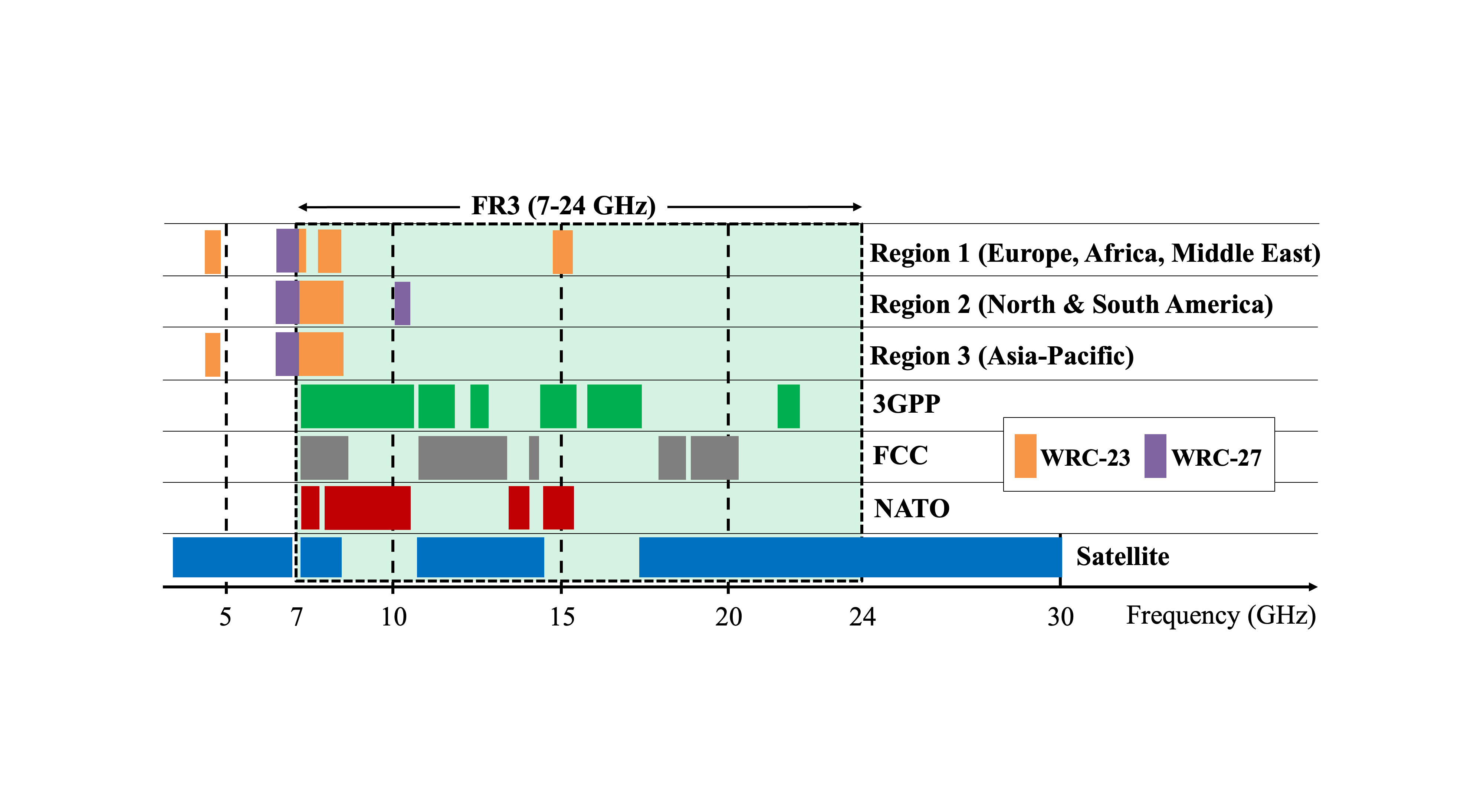}    
    \caption{Considered frequency allocations in \ac{fr3} for \ac{6g} and existing allocations for satellite and military usage. The use of non-contiguous bands in \ac{fr3} over a very wide bandwidth represents an opportunity but also a challenge for coherent multiband \ac{isac}.}
    \label{fig:fr3-bands}
\end{figure}

A further opportunity lies in the inherent disaggregation of \ac{fr3} spectrum. Allocations are fragmented across multiple non-contiguous bands due to incumbent services such as satellites, radio astronomy, and radar, as depicted in Fig.~\ref{fig:fr3-bands}. 
While this fragmentation poses challenges for traditional single-band systems, it naturally motivates multiband and frequency-agile designs. By dynamically hopping or aggregating carriers across the \ac{fr3} range, networks can trade-off coverage and rate depending on propagation and blockage conditions. 

Importantly, \ac{fr3}'s multiband nature also creates opportunities for \ac{isac}.
Rather than a drawback, the \ac{fr3} disaggregation can be turned into a feature. 
Coherent multiband \ac{isac} techniques can exploit non-contiguous \ac{fr3} bands to jointly enhance communication and sensing resolution by \textit{aggregating} the available bands, thus increasing the total bandwidth with minimal spectrum occupation.
However, several challenges arise in using such disaggregated wideband spectrum for coherent multiband \ac{isac}.

First, real targets exhibit frequency-dependent (\textit{incoherent}) electromagnetic scattering properties when signals having very different carrier frequencies are used.
Given the extremely wide span of \ac{fr3}, sensing targets can not be modeled as \textit{frequency isotropic}, i.e., idealized point-like scatterers that scatter the signal in the same way regardless of the carrier frequency, as commonly done in multiband \ac{isac} works~\cite{pegoraro2024hisac, li2025enabling}.
This means that the complex-valued coefficients of such targets in the \ac{cir} are not constant in frequency, which invalidates existing coherent multiband algorithms and phase offset compensation techniques across bands. 
The incoherence becomes more significant in \textit{high fractional bandwidth} scenarios, i.e., when the total bandwidth exceeds $20$~\% of the carrier frequency. 
Unlike FR1 (low carrier frequency, narrow bandwidth) and FR2 (high carrier frequency, wide bandwidth), \ac{fr3} can reach over $100$\% fractional bandwidth if multiple disaggregated bands are combined.

Second, the use of disaggregated spectrum introduces \textit{grating lobes} in the range profile obtained by aggregating multiple bands. 
This is because uniform sampling in the frequency domain, i.e., without gaps, is a necessary condition to obtain an impulse-like range profile for sensing targets. 
This significantly degrades the sensing results, introducing ambiguity in the targets' detection and nullifying the multiband resolution gain.

To date, existing research has focused on modeling and experimentally characterizing \ac{fr3} for communications and \ac{isac}~\cite{azim2025statistical} without considering the challenges of coherent multiband sensing.
Specifically: (i)~\textit{experimental characterization} of frequency anisotropy over \ac{fr3} band is lacking, so it is unclear when existing multiband algorithms are applicable and when instead they are not,  (ii)~\textit{metrics} to evaluate the \textit{phase coherence} of common targets and the impact of high fractional bandwidth on multiband \ac{isac} algorithms are missing, and (iii)~analysis and \textit{countermeasures for the grating lobes} due to non-contiguous bands are still in their early stages and no results are available for \ac{fr3} considered allocations yet.

In this work, we tackle the above challenges by presenting the first experimental characterization for coherent multiband \ac{isac} over \ac{fr3}, including novel coherence evaluation metrics and bandwidth aggregation algorithms. We focus on \textit{ranging}, i.e., the measurement of a target's distance from the \ac{isac} device, combining multiple frequency bands coherently.
This is commonly the first step in \ac{isac} systems, to localize the targets of interest, hence our results also apply to other, more advanced, sensing applications such as motion analysis and respiration monitoring, among others.
Our contributions are summarized below.
\begin{enumerate}
    \item We present the first experimental characterization of coherent multiband \ac{isac} over the \ac{fr3} band. Our analysis encompasses both the frequency anisotropy phenomenon and the impact of non-contiguous bands. The collected data is made available to the research community at \url{https://zenodo.org/uploads/17100726}.
    \item We propose three novel metrics to evaluate the phase coherence of common targets, including humans, to check the applicability of existing multiband \ac{isac} algorithms and assess their performance. 
    \item We propose a heuristic algorithm, named \ac{spbp}, for coherent multiband combination that mitigates the grating lobes due to the use of non-contiguous subbands and outperforms existing approaches like \ac{bp} and \ac{omp} on our data. 
\end{enumerate}
Our findings point out that the high fractional bandwidth of \ac{fr3} and the frequency anisotropy of common targets is a neglected but key aspect to take into account in the design of multiband \ac{isac} algorithms and their integration in \ac{6g}.
In addition, our results suggest that existing multiband aggregation algorithms are unlikely to be effective when considering very wide portions of \ac{fr3} and real extended targets such as humans. 

The paper is organized as follows. 
In \secref{sec:results}, we present the main results from our experimental characterization of phase coherence and fragmented spectrum, along with the proposed metrics.
In \secref{sec:discussion}, we discuss the main takeaways from the experimental results and we compare them to the existing literature.
In \secref{sec:methods}, we detail the methodology used in the paper, including the system and signal models, the description of the algorithms, and the experimental protocol.
Supplementary material with additional information on the methodology is provided as a separate file.

\section{Results}\label{sec:results}

In this section, we present our results on \ac{fr3} multiband measurements. 
For a detailed explanation of the system model and algorithms, the reader is referred to \secref{sec:methods}.

\noindent Our objectives are:
\begin{enumerate}
    \item To provide a practical way of evaluating the phase coherence of sensing targets and the applicability of existing multiband aggregation algorithms that assume frequency isotropic targets.
    
    \item To experimentally demonstrate and quantify the non-coherence of complex targets when a large bandwidth is aggregated (several GHz), providing insights into the difference between extended and point-like targets.
    
    \item To evaluate the impact of non-contiguous subbands allocation in \ac{fr3}, and propose countermeasures to mitigate the appearance of grating lobes.
    For this, we focus on \textit{the actual frequency bands} considered by \ac{3gpp} for \ac{6g}.
\end{enumerate}  
To this end, we collect a dataset of \ac{cfr} measurements in \ac{fr3} with different target types (corner reflector, metal plate, and humans), as detailed in \secref{sec:dataset}.
We then propose three novel metrics to evaluate phase coherence and use them to address objective $1$ (\secref{sec:metrics}).
Experimental results in terms of the three metrics are provided, achieving objective $2$ in \secref{sec:coherence-results}.
Finally, in \secref{sec:resolution-results}, we tackle objective $3$ by analyzing the grating lobes for actual \ac{fr3} subbands and comparing a newly proposed multiband aggregation algorithm to two state-of-the-art methods from the \ac{isac} literature.

\subsection{Dataset}\label{sec:dataset}

In our experimental campaign, we collect a dataset of $475$ \ac{ota} \ac{cfr} measurements obtained with a multiband monostatic transceiver spanning the frequency range $6$-$22$~GHz. 
The transceiver sweeps the full frequency range, transmitting \ac{ofdm} pilot signals with a bandwidth of $500$~MHz or $1$~GHz with different carrier frequencies. More details on the frequency sweep and the measurement device can be found in \secref{sec:setup}.
From each \ac{cfr} measurement, the corresponding \ac{cir} is obtained by applying an \ac{idft} along the \ac{ofdm} subcarriers dimension.
The dataset consists of $200$ calibration measurements, $150$ single-target measurements, and $125$ multitarget measurements. 
We consider three target types: a calibrated corner reflector, a metal plate, and humans, as detailed in \secref{sec:targets}.
To evaluate the impact of using non-contiguous subbands, in some of our experiments, we extract \textit{subsets} of the subbands collected in the frequency sweep that do not cover the full frequency range.
This allows us to generate a large number of non-contiguous subband patterns that we use to evaluate the considered algorithms in different scenarios.
As complementary data, we also collect phase response measurements for the corner reflector using a \ac{vna} over the bandwidth $6$-$24$~GHz. These measurements constitute a reference for the expected target phase response, since the \ac{vna} is calibrated and does not introduce phase distortions across the measured band. 

\subsection{Multiband combination algorithms}\label{sec:algorithm}

We consider two multiband \ac{isac} algorithms from the literature, \ac{bp} and \ac{omp}, and propose a third one, called \ac{spbp}, to mitigate the impact of grating lobes due to non-contiguous subbands.
\ac{bp} is used as the default algorithm in single-target measurements with contiguous subbands (\secref{sec:coherence-results}), since it is not affected by grating lobes in this case. The results obtained with \ac{omp} and \ac{spbp} are instead presented in \secref{sec:resolution-results} to evaluate the impact of grating lobes on multitarget resolution.
The algorithms are listed in the following, and explained more in depth in \secref{sec:mb-algorithms}.

\begin{itemize}
    \item \textbf{\ac{bp}}: This method is commonly used in the \ac{sar} literature and has been used in \ac{isac} in~\cite{tagliaferri_wavefield,tagliaferri2024cooperative}. It combines the subbands by oversampling the \ac{cir} of each subband, compensating for the propagation phase term of each \ac{cir}, and summing the \acp{cir} coherently.
    Grating lobes due to non-contiguous subbands affect the result since no mitigation is performed.
    
    \item \textbf{\ac{omp}}: This algorithm is based on \ac{cs}~\cite{eldar2012compressed} and it was previously used in multiband sensing in~\cite{pegoraro2024hisac}. It tackles the combination of the subbands as a sparse reconstruction problem. It is well-suited to mitigate the grating lobes due to non-contiguous subbands. 
    However, the nominal multitarget resolution of \ac{omp} is hard to estimate and certainly lower than that of \ac{bp}.
    Moreover, \ac{omp} is an \textit{on-grid} method, meaning its performance heavily depends on the granularity of a pre-defined grid of possible range values. 
    For a detailed description of the \ac{omp} algorithm applied to our measurements, we refer to the supplementary material (Supplementary note - 2).
    
    \item \textbf{\ac{spbp}}: This method is based on \ac{bp} but attempts to mitigate the grating lobes by using non-linear processing on the multiband range profile (see \secref{sec:subsets-prod-bp}).
    Specifically, \ac{spbp} splits the available subbands into two sets, optimized to have grating lobes at different locations, then applies \ac{bp} to each of them and takes the product of the resulting combined \acp{cir}. As a result, grating lobes are strongly attenuated by the product while the targets persist. 
\end{itemize}

\subsection{Performance metrics}\label{sec:metrics}

In this section, we introduce the metrics used in the experimental results to evaluate the coherence of targets in multiband \ac{isac} and the multitarget ranging resolution and accuracy in empirical settings.

\subsubsection{Multiband target coherence metrics}
Assessing the multiband coherence, i.e., coherence over a wide range of frequencies, of an extended target is challenging and requires considering the \acp{cir} obtained in different bands and the combined \textit{range profile} as a result of multiband processing. 
The range profile contains the complex-valued coefficients of scattering points in the environment as a function of the distance (range) from the \ac{isac} device.
Coefficients with high magnitude indicate the presence of scatterers or reflectors at certain ranges.

Our contribution is to propose novel metrics to evaluate: \textit{(i)}~the phase coherence of the target in the different measured subbands and \textit{(ii)}~the quality of the resulting multiband range profile. 
We design three metrics, which are a function of the target index $\ell$, detailed in the following.
In the definitions below, we denote by $R_{\ell}$ the range value corresponding to target $\ell$ and by $\eta_k(R)$ the range profile in subband $k$ evaluated at distance $R$. We consider a total of $K$ combined subbands. 
The range profile is obtained from the \ac{cir} of subband $k$ by compensating for the contribution of the carrier phase, as explained in detail in~\secref{sec:bp-raf}, \eq{eq:prop-phase-comp}.
\begin{enumerate}
    \item \textbf{\ac{mpc}}.
    The \ac{mpc} measures the spread of the scattering phases across the subbands as
    \begin{equation}\label{eq:mpc}
        \mathrm{MPC}(\ell) =  \left|\frac{1}{K}\sum_{k=0}^{K-1} e^{j \angle{\eta_k(R_{\ell}) }}\right| ,
    \end{equation}
    where $\angle{\cdot}$ is the phase operator.
The \ac{mpc} lies in $[0, 1]$ and equals $1$ if the phases are all identical across the subbands, indicating maximal coherence. In this case, the terms $e^{j\angle{\eta_k(R_\ell})}$ all add up coherently. Conversely, its value is $0$ if the phases combine destructively, leading to a perfect cancellation of the $e^{j\angle{\eta_k(R_\ell)}}$ in the sum, indicating maximum incoherence. 
    The target locations $R_\ell$ to be used in \eq{eq:mpc} are obtained by performing peak detection on the range profiles obtained in multiple subbands.
    The \ac{mpc} is very informative since it directly evaluates the phase coherence of the $K$ multiband range profiles when considering a specific target. However, it only \textit{indirectly} evaluates the quality of the combined multiband range profile. Indeed, it does not capture slight misalignments of the range profile peaks in the different subbands, which may reduce the quality of the final combination.
    
    \item \textbf{\ac{nmpm}.} The \ac{nmpm} evaluates the relative magnitude of the peak corresponding to a target in the combined range profile compared to the average magnitude of the same target across the single subbands. Denoting the multiband combined range profile as $\eta(R)$, the \ac{nmpm} in dB is
    \begin{equation}
        \mathrm{NMPM}(\ell) = 20 \log_{10} \left(\frac{|\eta(R_{\ell})|}{\frac{1}{K}\sum_{k=0}^{K-1} |\eta_k(R_\ell)|}\right).
    \end{equation}
    If, after multiband combination, the magnitude of the target peak lowers with respect to the mean peak magnitude across subbands, this is a sign that the combined subbands are not coherent. 
    This could be either due to phase incoherence, as evaluated by the \ac{mpc}, or due to slight peak misalignment or other inconsistencies in the range profiles between the subbands. 
    Hence, \ac{nmpm} evaluates the quality of the range profiles combination rather than the multiband coherence directly.

    \item \textbf{\ac{empw}.} This additional metric evaluates the empirical $3$ dB peak width in the multiband combined range profile. This is a direct measure (in meters) of the resolution of the range profile after multiband processing, and can be compared to the theoretical resolution given by $c/(2B)$ where $c$ is the speed of light and $B$ is the bandwidth of the signal (either the single subband, $B_k$, or the total multiband occupation). 
    The \ac{empw} is obtained as 
    \begin{equation}
        \mathrm{EMPW}(\ell) = \mathrm{supp}_{- 3 \mathrm{dB}} |\eta(R)| = \left\{R \,|\, \eta(R) \geq \eta(R_{\ell})/2 \right\},
    \end{equation}
    where $\mathrm{supp}_{- 3 \mathrm{dB}}$ denotes the $-3$~dB support of the range profile around target $\ell$. 
Similar to \ac{nmpm}, \ac{empw} evaluates the quality of the multiband range profile and is an indirect measure of the multiband coherence. 
\end{enumerate}

\subsubsection{Empirical multitarget resolution and accuracy} 
To evaluate the empirical multitarget resolution when we aggregate different subbands, we use the \ac{ospa} metric from the literature~\cite{schumacher2008consistent}. 
The \ac{ospa} measures the distance between sets of points, effectively capturing the contribution of cardinality differences and distance between the points in the sets. It is widely used in the evaluation of multitarget tracking algorithms. 
Intuitively, the \ac{ospa} finds the minimum average Euclidean distance between the points in the two sets over all possible permutations of the points, then it adds a penalty of $\mu$ for each unit difference in the cardinality of the two sets.
In this computation of the Euclidean distance, the distance values are cropped at a maximum value of $\mu$, so that the \ac{ospa} values are in $[0, \mu]$.
Further details on the formulation of the \ac{ospa} metric are given in the supplementary material (Supplementary note - 1).

\subsection{Results for multiband target coherence}\label{sec:coherence-results}

In this section, we present our experimental results on the coherence of single targets in wide multiband settings in \ac{fr3}. We start by providing qualitative results of the range profile obtained by combining multiple bands across different bandwidths and carrier frequencies. Then, we give a quantitative evaluation of the targets' coherence in terms of the metrics proposed in~\secref{sec:metrics}.

\subsubsection{Corner reflector multiband range profiles}
\newlength{\globaltextwidth}
\setlength{\globaltextwidth}{\textwidth}

\begin{figure*}[t!]
	\begin{center}   
		\centering
        \includegraphics[width=\textwidth]{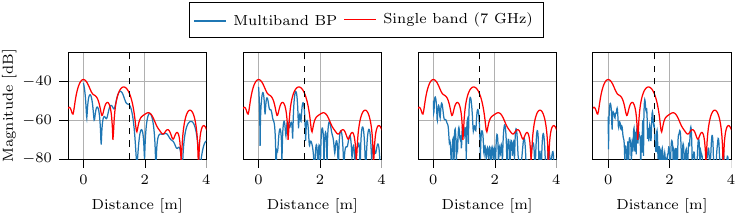} 
        \caption{Qualitative plots of the multiband range profile obtained using \ac{bp} with a \textbf{corner reflector}, at $6.5$ GHz starting frequency. Different plots from left to right represent the combination of 1, 3, 5, and 7 GHz of bandwidth. The black dashed line represents the location of the target measured with a laser telemeter.}
		\label{fig:cir-example-6.5}
	\end{center}
 \vspace{-0.5cm}
\end{figure*}

In \fig{fig:cir-example-6.5}, we show the \ac{cir} obtained from a corner reflector target at $1.5$~m from the measurement device (vertical dashed line) with a $500$~MHz bandwidth and $6.5$~GHz carrier frequency (red curve), compared to the multiband range profile with \textit{contiguous subbands} using \ac{bp} (blue curve). 
To obtain the latter, we use the \ac{bp} algorithm detailed in~\secref{sec:bp-raf}. 
In the subfigures (a-d), we show the impact of changing the total combined bandwidth for multiband processing, which we call $B_{\rm tot}$, ranging from $1$ to $7$~GHz. $B_{\rm tot}$ is incremented by combining adjacent frequency bands of $500$~MHz starting from $6.5$~GHz. 
We notice the resolution improvement demonstrated by the significant reduction in the width of the main range profile peak corresponding to the target. When a bandwidth of $3$~GHz or more is coherently combined, this reveals a \textit{secondary peak}, $6$~dB below the main one, that is not observed using $500$~MHz bandwidth. 
This is likely due to a multipath reflection.
Increasing the total combined bandwidth, we also observe a reduction in the peak height of the main target, with an overall loss of around $3$~dB at $B_{\rm tot} = 7$~GHz.
The reduction is due to the imperfect target coherence across frequencies and hardware non-ideality, as further investigated in the next section.
In the case of a perfectly coherent combination of different subbands and ideal hardware, the peak should not decrease in magnitude, since all the target's contributions would sum in phase. 
Note that the strong peak around $0$~m is due to the direct signal path from the transmitter to the receiver antenna.

\begin{figure*}[t!]
	\begin{center}   
		\centering
        \includegraphics[width=\textwidth]{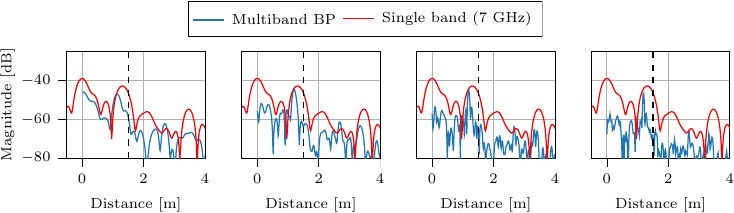} 
		\caption{Qualitative plots of the multiband range profile obtained using \ac{bp} with a \textbf{corner reflector}, at $14$ GHz starting frequency. Different plots from left to right represent the combination of 1, 3, 5, and 7 GHz of bandwidth. The black dashed line represents the location of the target measured with a laser telemeter.}
		\label{fig:cir-example-14}
	\end{center}
 \vspace{-0.5cm}
\end{figure*}

In \fig{fig:cir-example-14}, the same experiment is repeated, starting from $14$~GHz. 
We notice that the multiband range profile exhibits a main target peak located at the same distance as in the $6.5$~GHz experiment of \fig{fig:cir-example-6.5}, which confirms the target presence. 
A similar reduction of $3$~dB in main peak magnitude is evident, going from $1$ to $7$~GHz total bandwidth.
However, the secondary peak observed in \fig{fig:cir-example-6.5} is not present, demonstrating that besides the total combined bandwidth, another critical parameter to consider in multiband \ac{isac} is the carrier frequency.
The different scattering response at different frequencies could be due to the size of the corner reflector compared to the wavelength: reducing the wavelength makes the size of the corner relatively larger from an electromagnetic perspective, reducing border effects.

This experiment suggests that exploring the variations of the target coherence metrics across different values of $B_{\rm tot}$ and carrier frequency is necessary to gain a deeper insight into how to design multiband sensing algorithms.

\subsubsection{Human multiband range profiles}
\begin{figure*}[t!]
	\begin{center}   
		\centering
        \includegraphics[width=\textwidth]{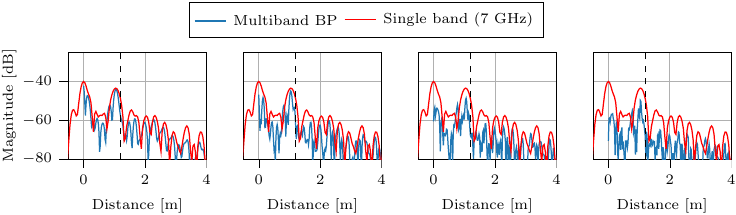} 
		\caption{Qualitative plots of the multiband range profile obtained using \ac{bp} with a \textbf{human}, at $6.5$ GHz starting frequency. Different plots from left to right represent the combination of 1, 3, 5, and 7 GHz of bandwidth. The black dashed line represents the location of the target measured with a laser telemeter.}
		\label{fig:cir-example-human-6.5}
	\end{center}
 \vspace{-0.5cm}
\end{figure*}

\begin{figure*}[t!]
	\begin{center}   
		\centering
        \includegraphics[width=\textwidth]{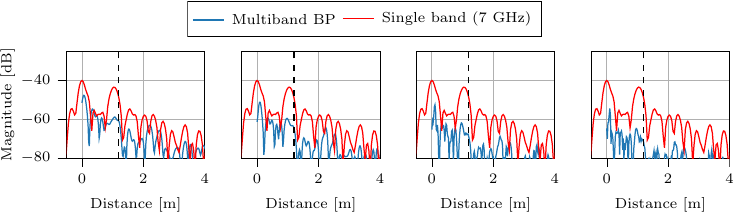} 
		\caption{Qualitative plots of the multiband range profile obtained using \ac{bp} with a \textbf{human}, at $14.5$ GHz starting frequency. Different plots from left to right represent the combination of 1, 3, 5, and 7 GHz of bandwidth. The black dashed line represents the location of the target measured with a laser telemeter.}
		\label{fig:cir-example-human-14}
	\end{center}
 \vspace{-0.5cm}
\end{figure*}

In \fig{fig:cir-example-human-6.5} and \fig{fig:cir-example-human-14}, we repeat the experiments of the previous section with a human target instead of a corner reflector. 
To reduce the total frequency sweep duration and to avoid artifacts due to the person's involuntary movements, we increase the single band bandwidth to $1$~GHz. We notice that the results obtained starting from $6.5$~GHz (\fig{fig:cir-example-human-6.5}) or $14.5$~GHz (\fig{fig:cir-example-human-14}) are now completely different. 
Starting from $6.5$~GHz, the range profile presents a clear target peak, which narrows as $B_{\rm tot}$ increases. 
However, a secondary peak with slightly lower magnitude appears when increasing the combined bandwidth above $1$ GHz. 
This represents a secondary scattering center of the target, due to the more complex structure of the human body with respect to the corner reflector.
Conversely, at $14.5$~GHz the peak corresponding to the target is barely visible and does not seem to gain significant resolution when $B_{\rm tot}$ increases.

This experiment demonstrates that for complex extended targets: \textit{(i)} the target response in the range profile deviates from a single clear peak when multiple target parts combine coherently, especially when combining multiple GHz of bandwidth over a large \textit{fractional bandwidth}, \textit{(ii)} the number of visible scattering centers highly depends on the carrier frequency of the total combined bandwidth (frequency anisotropy).

\subsubsection{Quantitative evaluation of the multiband target coherence}\label{sec:coherence-quant}
\begin{figure*}[t!]
    \centering
    \includegraphics[width=\textwidth]{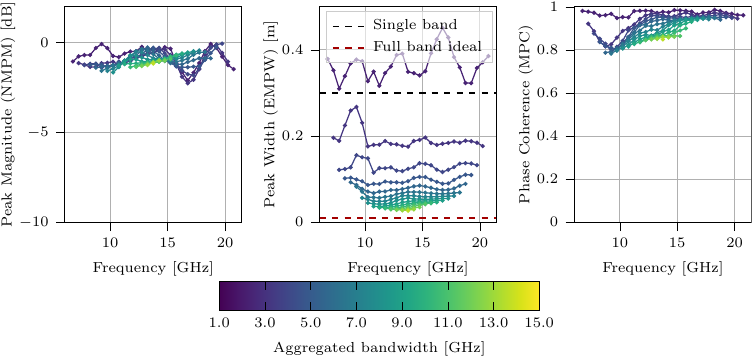}
    \caption{Coherence metrics of a corner reflector target in the range $6$-$22$~GHz. The bandwidth of a single subband is $0.5$~GHz. We report the \ac{nmpm}, \ac{empw}, and the \ac{mpc} obtained for different carrier frequencies (on the $x$-axis) and different total combined bandwidth (corresponding to the different colored curves as specified in the colorbar). For the \ac{empw} we report the single band and full band ideal resolutions with dashed black and red lines, respectively. }
    \label{fig:corner-coherence}
\end{figure*}

\begin{figure*}[t!]
    \centering
    \includegraphics[width=\textwidth]{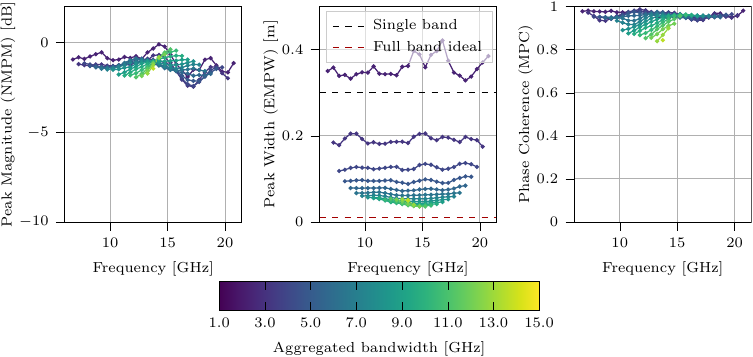} 
    \caption{Coherence metrics of a metal plate target in the range $6$-$22$~GHz. The bandwidth of a single subband is $0.5$~GHz. We report the \ac{nmpm}, \ac{empw}, and the \ac{mpc} obtained for different carrier frequencies (on the $x$-axis) and different total combined bandwidth (corresponding to the different colored curves as specified in the colorbar). For the \ac{empw} we report the single band and full band ideal resolutions with dashed black and red lines, respectively. }
    \label{fig:metalplate-coherence}
\end{figure*}

\begin{figure*}[t!]
    \centering
   \includegraphics[width=\textwidth]{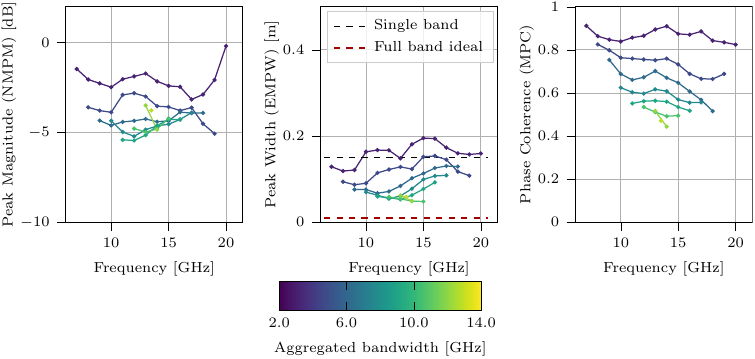} 
    \caption{Coherence metrics of a human target in the range $6$-$22$~GHz. The bandwidth of a single subband is $1$~GHz. We report the \ac{nmpm}, \ac{empw}, and the \ac{mpc} obtained for different carrier frequencies (indicated on the $x$-axis) and different total combined bandwidth (corresponding to the different colored curves as specified in the colorbar). For the \ac{empw} we also report the single band and full band ideal resolutions with dashed black and red lines, respectively.}
    \label{fig:human-coherence2}
\end{figure*}

In \fig{fig:corner-coherence}, \fig{fig:metalplate-coherence}, and \fig{fig:human-coherence2}, we show our results for multiband target coherence using the three targets of \secref{sec:targets}, respectively. For each target, we show the three metrics proposed in \secref{sec:metrics} for different values of $B_{\rm tot}$, which is the contiguous aggregated bandwidth, and carrier frequency.
Specifically, each curve (with a different color) represents the result obtained by combining a different $B_{\rm tot}$. 
Darker colors represent narrower $B_{\rm tot}$ while lighter colors represent wider $B_{\rm tot}$.
Each point in the curves is obtained with a different value of the carrier frequency, as specified by the $x$-axis.
Note that curves corresponding to wider $B_{\rm tot}$ also contain fewer points, since there are fewer possibilities to fit a wide bandwidth in the considered frequency range of $6-22$~GHz.
All the results shown in the following are obtained by averaging $50$ measurements taken in a row, with a few seconds' pause between subsequent captures. 
A moving average smoothing filter along the frequency dimension, with a window of $3$ points, is applied to the averaged curves with more than $5$ points to improve readability.

\textbf{Corner reflector:} \fig{fig:corner-coherence} shows that the corner reflector can be considered coherent over a very wide bandwidth, with slight differences depending on the carrier frequency.
The phase coherence (\ac{mpc} metric) is above $0.8$ for almost all the considered carrier frequencies and bandwidth values. 
It shows its minimum for a carrier frequency of $9$ GHz, consistently across all values of $B_{\rm tot}$. Additionally, it gradually decreases as $B_{\rm tot}$ increases. It stabilizes to $0.85$, which is a relatively high coherence, when combining the full frequency range $6-22$~GHz.
The main range profile peak magnitude loss with respect to the mean magnitude of the combined subbands (\ac{nmpm} metric) does not fall below $-2$~dB regardless of the carrier frequency, and shows the most significant degradation ($1.9$~dB) when combining $1$ to $4$~GHz of bandwidth around a carrier frequency of~$17$~GHz.
The \ac{empw} results are coherent with the other metrics, showing that the main peak width of the target consistently shrinks when the combined bandwidth increases. 
The final resolution approaches the ideal one (red dashed line), but does not achieve it, likely due to the combined effect of imperfect target coherence (lower than $1$), antenna response non-ideality, and the residual impact of noise on the channel estimation process.

\begin{figure}[t!]
	\begin{center}   
		\centering
\includegraphics[width=0.7\columnwidth]{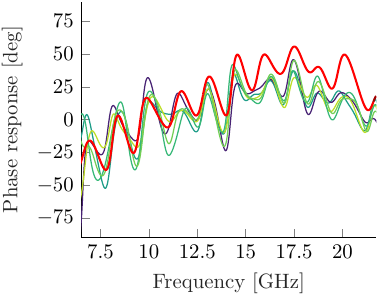} 
		\caption{\ac{vna} phase response for a corner reflector (colors blue-yellow indicate $5$ different trials) compared to the response of the antenna measured on a point-to-point link (red line).}
		\label{fig:corner-vna}
	\end{center}
 \vspace{-0.5cm}
\end{figure}

In \fig{fig:corner-vna}, we provide the phase response obtained from a \ac{vna} measurement on the corner reflector to verify that it can be considered coherent over a wide bandwidth.
We report the phase response over the full bandwidth obtained in $5$ trials (blue to yellow curves) compared to the phase response obtained from a point-to-point link (red curve) without any target to measure the phase response of the antenna. 
Our results show that the phase presents oscillations of $\pm 25^{\circ}$ and a general increasing trend. 
However, both these variations have to be mainly attributed to the non-ideal antenna response, which shows very similar variations.
This shows that, in terms of our metrics, the $\pm 25^{\circ}$ phase oscillations due to the antenna response do not have a significant impact on the coherence of the target and still allow coherent multiband processing over several GHz of bandwidth.
Additional information on how we collected the measurements to obtain \fig{fig:corner-coherence} and the \ac{vna} \ac{cir} is provided in the supplementary material (Supplementary note - 4).

\textbf{Metal plate:} 
\fig{fig:metalplate-coherence} shows that the metal plate target has a similar behavior to the corner reflector, showing high coherence over a very wide bandwidth.
The \ac{mpc} is above $0.8$ for all the considered carrier frequencies and bandwidth values. 
Similarly to the corner reflector, the target appears to be slightly less coherent when lower carrier frequencies are considered, i.e., less than $14$~GHz. 
This may be because at higher carrier frequencies, the dimensions of the corner reflector and metal plate become large compared to the wavelength, reducing border effects and non-idealities.
Another possibility is that the slight incoherence for frequencies lower than $14$~GHz is due to the flatter phase response of the antenna in the range $14$-$22$~GHz (see \fig{fig:corner-vna}).
The coherence obtained by combining $15$~GHz of bandwidth is slightly lower than in the corner reflector case.
The \ac{nmpm} is also quite constant and shows the most significant degradation ($2.1$~dB) when combining $1$ to $4$~GHz of bandwidth around a carrier frequency of~$17$~GHz. 
As in the corner reflector case, since this degradation at $17$ GHz is not evident in the \ac{mpc}, it is likely due to the antenna response or to a residual effect of the measurement device rather than to a property of the target.
The \ac{empw} shows that the main peak width of the target consistently shrinks when the combined bandwidth increases, except for very wide bandwidth values above $13$~GHz. 
In this case, the resolution stops improving and slightly degrades, increasing the peak width. This can be appreciated in \fig{fig:metalplate-coherence} by observing that the \ac{empw} corresponding to $B_{\rm tot}=15$ GHz is higher than its $B_{\rm tot}=12$ GHz counterpart, meaning that the effective range resolution of the system degrades.
This is in line with the phase coherence degradation observed for very wide combined bandwidths. 

\textbf{Human:} 
In \fig{fig:human-coherence2}, we report our results for the human target, which include measurements with a person at $1$~m and $1.2$~m from the measurement device, performing frequency sweeps with $1$~GHz bandwidth for a single subband.
In this case, \ac{mpc} shows a decreasing trend with an increasing carrier frequency. 
This trend is consistent across multiple values of the combined bandwidth.
A bandwidth of $2$~GHz can be combined with a moderate decrease in the coherence, which is over $0.8$ for all carrier frequencies. 
Conversely, combining bandwidths over $3$~GHz significantly degrades the phase coherence down to lower than $0.5$.
This indicates that a human does not provide a coherent point reflector-like response when increasing the resolution to the cm-level, differently from the corner reflector or metal plate targets.
Although the time taken for a frequency sweep is at most $150$~ms in this case (see \secref{sec:freq-sweep}), for high carrier frequencies and wide bandwidth, minimal movements of the subjects could also contribute to slightly degrade coherence.
The \ac{nmpm} and \ac{empw} metrics are in agreement with the \ac{mpc}.
The \ac{empw} falls to $-5$~dB around $17-18$~GHz carrier frequency and shows an irregular behavior compared to the corner reflector and metal plate cases.
The \ac{empw} shows that the resolution gain for the human target is higher at lower carrier frequencies, which is consistent with the lower coherence of the subbands for higher carrier frequencies observed in the \ac{mpc} and \ac{nmpm}.
This could be because at higher frequencies the wavelength of the signal interacts differently with (i)~body tissues due to the higher penetration depth compared to conducting materials, and (ii)~the roughness of the targets, including clothes and different body parts.
In complex targets such as humans, this produces an unpredictable effect on the scattering phase. 

\textit{Remark:} We verified that changes with respect to the corner reflector and metal plate results are due to the different target and not to hardware artifacts due to the wider employed bandwidth ($1$~GHz compared to $0.5$~GHz).
Specifically, the average \ac{mpc} obtained with $1$~GHz subbands with the corner reflector and $15$~GHz combined bandwidth (worst case) is $0.77$, while the \ac{nmpm} and \ac{empw} are $-2$~dB and $2.5$~cm, respectively. These values are comparable to those obtained with $0.5$~GHz subbands and indicate significantly higher coherence than those obtained with a human target.

\subsection{Results for multitarget resolution}\label{sec:resolution-results}

In this section, we evaluate the multitarget resolution capabilities of multiband processing with non-contiguous subbands.
We present experimental results using spectrum portions being considered for \ac{fr3} frequency allocations by \ac{3gpp} \cite{bazzi2025upper, testolina2024sharing}, discussed at the \ac{wrc} in $2023$. 
Specifically, we consider the spectrum portions  $7.125-10.5$~GHz, $12.7-13.25$~GHz, and $14.8-17.35$~GHz, divided into $5$ subbands determined by the frequency intervals $S_1 = [7.125, 8.5]$~GHz, $ S_2= [8.5, 10.5]$~GHz, $S_3 = [12.7, 13.25]$~GHz, $S_4 = [14.8, 15.35]$~GHz, and $S_5 = [15.35, 17.3]$~GHz.
Our measurement system collects \ac{cfr} measurements with $0.5$ or $1$~GHz bandwidth granularity, depending on whether we consider object targets (corner reflector or metal plate) or human targets, respectively.
Therefore, in the following experiments, we approximate the \ac{3gpp} intervals with sets of subbands collected by our system that include them, but do not match exactly the \ac{3gpp} bands.
We verified that this approximation does not significantly change the resulting \ac{raf}, which we analyze in the next section.

\textbf{\ac{raf} analysis:} 
The \ac{raf} represents the response of the multiband \ac{isac} systems to a point-like ideal reflector and highly depends on the location of the available subbands in the frequency domain and on their bandwidth.
We thoroughly discuss how to compute it in \secref{sec:bp-raf}, for different cases. 
In this case, we use the \ac{raf} expression of \eq{eq:bp-output-iso}, assuming frequency isotropic targets.

\begin{figure*}[t!]
	\begin{center}   
		\centering
		\subcaptionbox{Multiband \ac{raf} (left), and zoom on the main peak (right).\label{fig:raf}}[6cm]{\includegraphics[width=6cm]{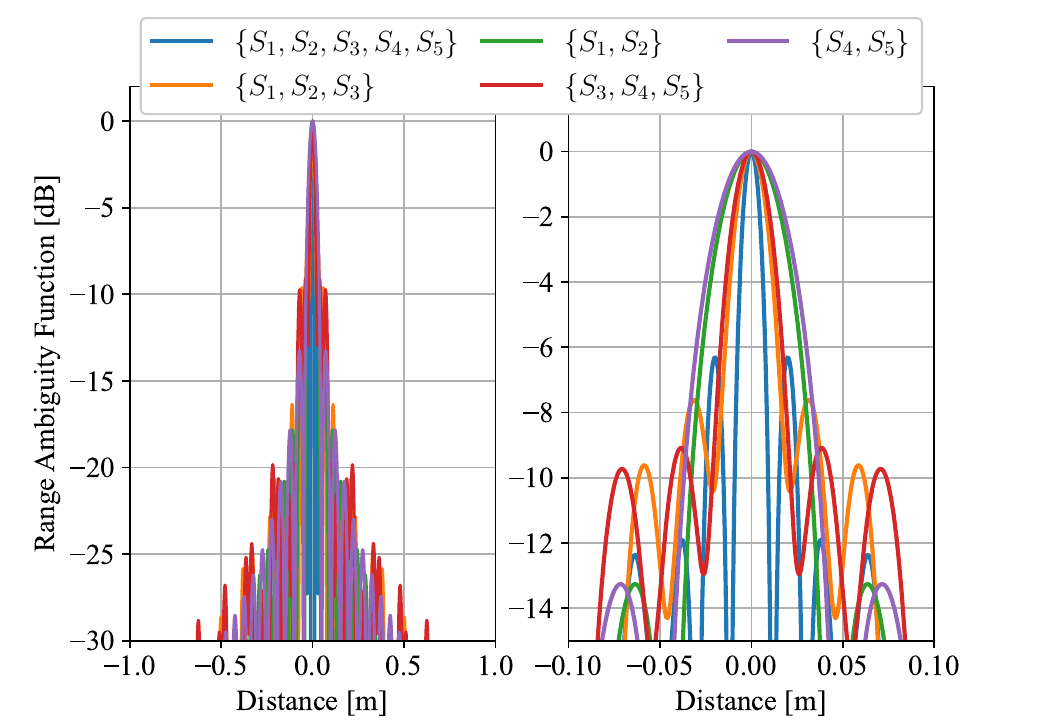}}
		\subcaptionbox{\ac{3gpp} \ac{fr3} considered frequencies (top) and our considered subbands in case of human targets (bottom).\label{fig:allocations}}[6cm]{\includegraphics[width=6cm]{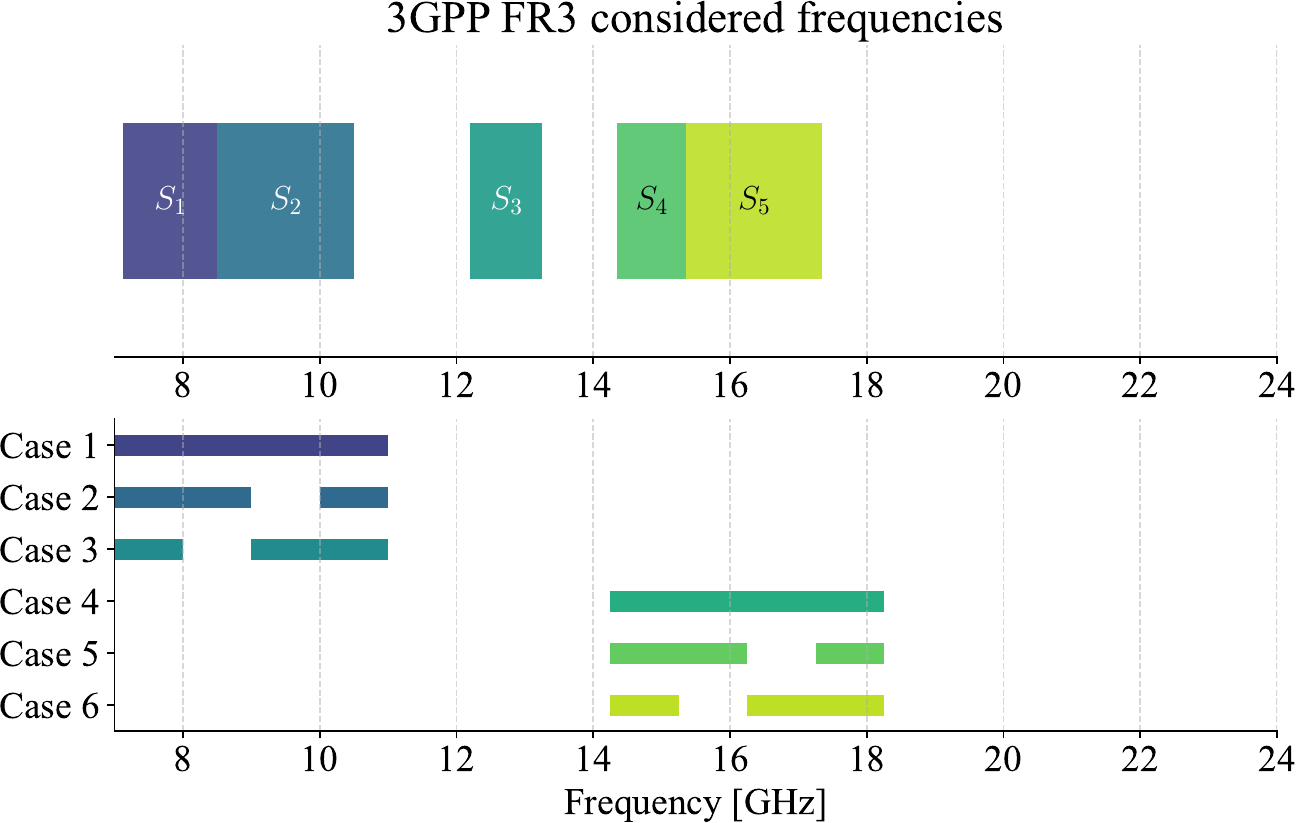}}
        \caption{\ac{raf} of different combinations of the multiband sets considered by \ac{3gpp} (a), \ac{3gpp} subbands and different sub-allocations considered in our experiments (b).}
		\label{fig:raf-3gpp}
	\end{center}
 \vspace{-0.5cm}
\end{figure*}

In \fig{fig:raf-3gpp}, we represent the ideal \ac{raf} obtained by combining different combinations of the $5$ subbands.
The aggregation of such a wide bandwidth with algorithms that assume a frequency isotropic target only makes sense for coherent targets like corner reflectors or metal plates, as discussed in~\secref{sec:coherence-results}.

Using $S_1$ and $S_2$ gives over $3$~GHz of \textit{contiguous} bandwidth, which leads to a resolution of $c/(2|S_1 \cup S_2|) = 4.4$~cm.
Similarly, the combination of $S_4$ and $S_5$ also provides a wide contiguous frequency band with a resolution of $c/(2|S_4 \cup S_5|) = 4.9$~cm.
Aggregating different combinations of the $5$ subbands improves the resolution at the cost of introducing grating lobes due to the non-contiguity of the subbands. 
Notably, aggregating $S_2$ and $S_3$ causes high grating lobes spaced by around $7$ cm that make this combination unsuitable for accurate ranging with multiple targets. 
Aggregating all the $5$ subbands gives the highest nominal resolution of $1.46$~cm, but the \ac{raf} exhibits a pair of grating lobes close to the main peak around $6$~dB below the main peak.
This reduces the nominal resolution in practice, especially in the case of close targets with very different \acp{rcs}.
Combining $S_1, S_2, S_3$ or $S_3, S_4, S_5$ lowers the grating lobes at the cost of a worse resolution (around $2.5$~cm). 
Notably, in this case, using $S_3, S_4, S_5$ is preferable since it leads to a narrower frequency gap and, in turn, $2.5$~dB lower grating lobes.
However, it may still be insufficient to properly resolve targets in practice if no compensation for the grating lobes is applied.

\begin{figure*}[t!]
	\begin{center}   
		\centering
		\subcaptionbox{Contiguous bandwidth using $S_1 $ and $ S_2$.\label{fig:cir-multi-corner-cont}}[4.25cm]{\includegraphics[width=4.25cm]{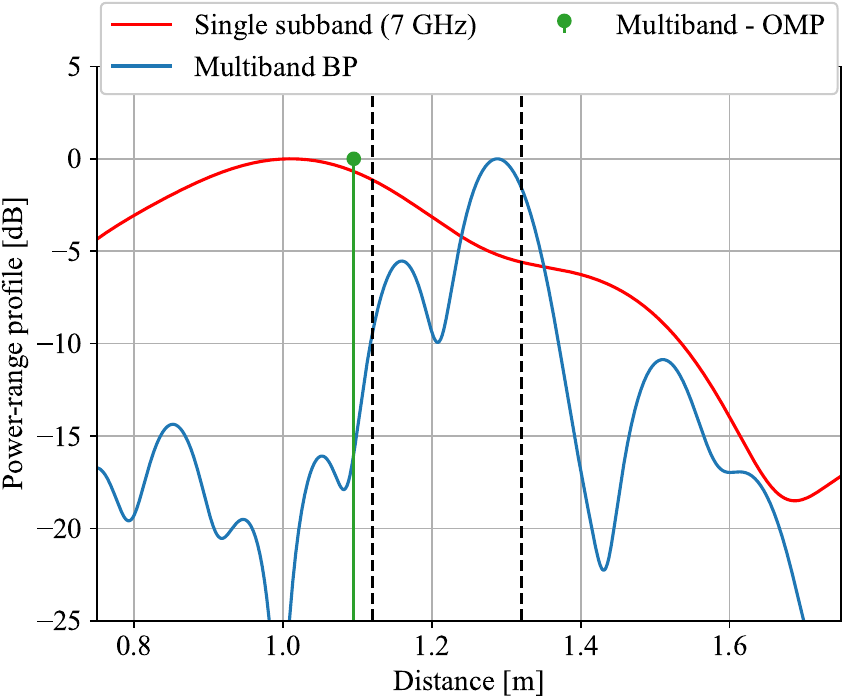}}
		\subcaptionbox{Non-contiguous bandwidth using $S_1, \dots, S_5$.\label{fig:cir-multi-corner-gap}}[4.25cm]{\includegraphics[width=4.25cm]{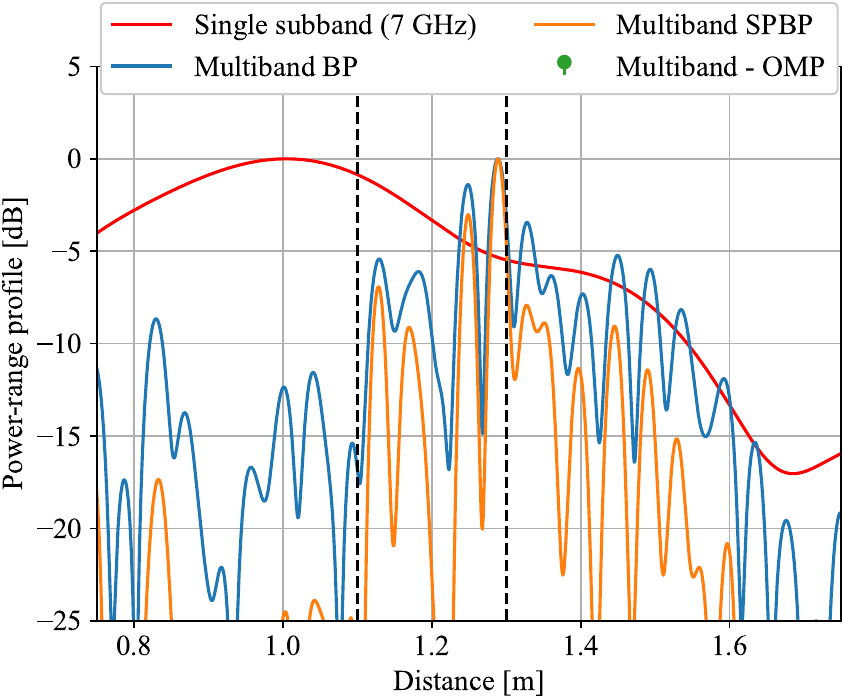}}
        \subcaptionbox{Contiguous bandwidth using $S_1 $,$ S_2$, and $S_3$.\label{fig:cir-multi-corner-cont-small}}[4.25cm]{\includegraphics[width=4.25cm]{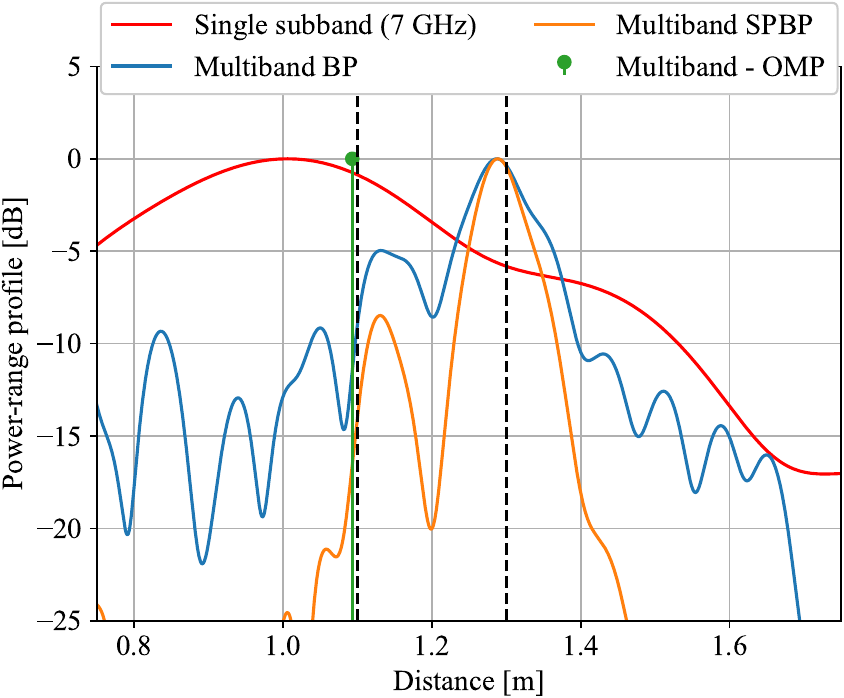}}
        \caption{Qualitative example of the combination of multiple subbands using \ac{bp}, \ac{omp}, and \ac{spbp} on corner reflector and metal plate targets with contiguous subbands (a), strongly non-contiguous subbands (b), and moderately non-contiguous subbands. Black dashed lines represent the measured real location of the targets.}
		\label{fig:cir-multi-corner}
	\end{center}
 \vspace{-0.5cm}
\end{figure*}

\textbf{Multiband range profile plots:} 
In \fig{fig:cir-multi-corner}, we report the multiband range profile squared magnitude obtained by \ac{bp}, \ac{omp}, and \ac{spbp} with two targets: a corner reflector and a metal plate.
In \fig{fig:cir-multi-corner-cont}, we use a contiguous bandwidth given by $S_1 \cup S_2$, so \ac{spbp} is not used since grating lobes are absent.
While \ac{omp} only detects one target, \ac{bp} correctly resolves both and matches their real location with an error of a few centimeters.
However, as shown in \fig{fig:cir-multi-corner-cont}, when using a non-contiguous bandwidth including all the subbands $S_1, \dots, S_5$, the grating lobes significantly degrade the multiband range profile. 
\ac{omp} does not find either of the two targets due to the aggregated effect of multiple slight incoherencies in the phases of the subbands.
\ac{bp} and \ac{spbp} exhibit significant grating lobes, and identifying the two targets is infeasible from the resulting multiband range profile.
In \fig{fig:cir-multi-corner-cont-small} we show the range profile obtained using non-contiguous subbands $S_1, S_2$ and $S_3$, which occupy a less fragmented spectrum portion than the previous case.
\ac{bp} can identify the two targets but is affected by significant grating lobes that reach up to $-5$~dB from the weaker target.
\ac{spbp} instead lowers the grating lobes down to $-15$~dB lower than the weaker target, improving the overall quality of the range profile. 
The cost for this improvement is a relative reduction of the magnitude of the weaker target with respect to the stronger one.

\begin{figure*}[t!]
	\begin{center}   
		\centering
		\subcaptionbox{Contiguous bandwidth  $7$-$11$~GHz (case $1$). \ac{bp} resolves the two subjects.\label{fig:cir-human-cont}}[5.cm]{\includegraphics[width=4.5cm]{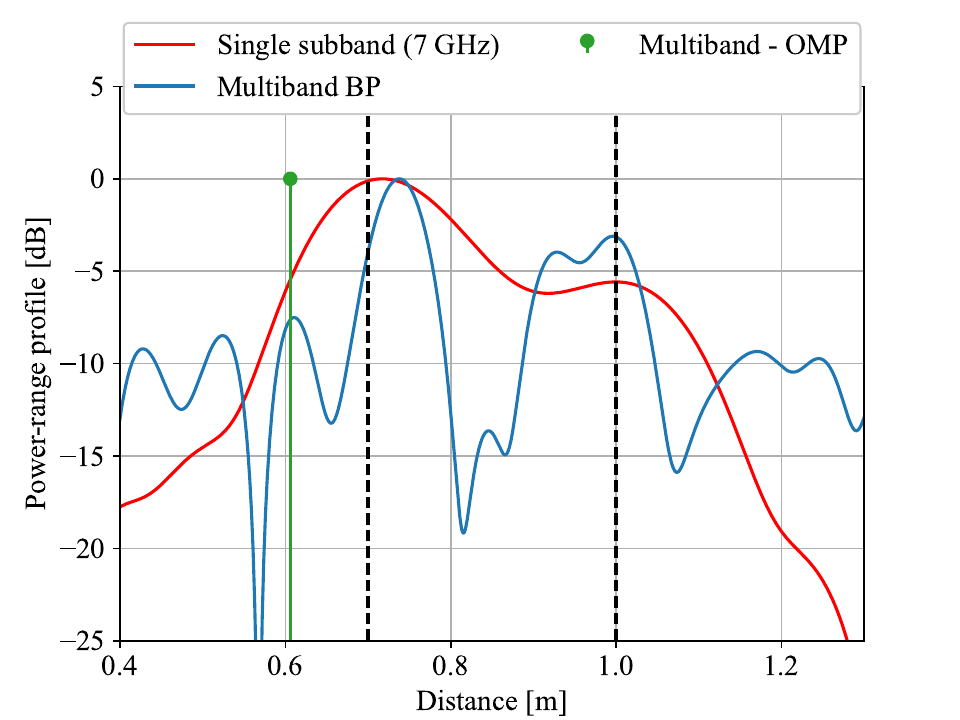}}
		\subcaptionbox{Gapped bandwidth $7$-$11$~GHz (case $3$). \ac{spbp} mitigates the grating lobes due to non-contiguous subbands.\label{fig:cir-human-gap}}[5.cm]{\includegraphics[width=4.5cm]{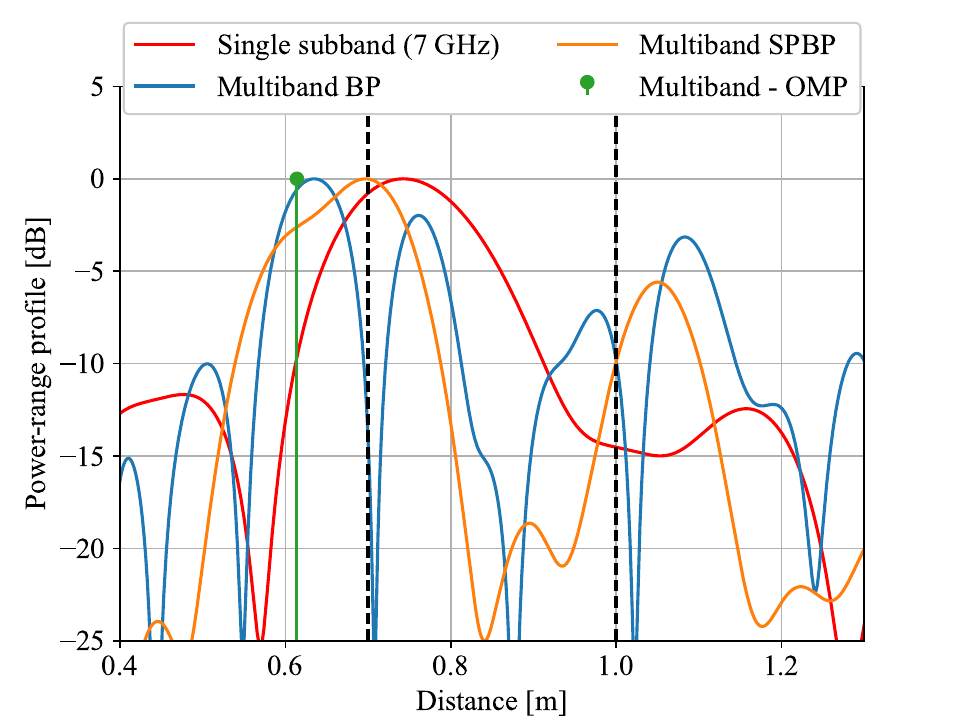}}
        \caption{Qualitative example of the range profile squared magnitude obtained with the combination of multiple subbands using \ac{bp}, \ac{omp}, and \ac{spbp} on two human targets with contiguous (a) and non-contiguous (b) subbands.}
		\label{fig:cir-multi-human}
	\end{center}
 \vspace{-0.5cm}
\end{figure*}

In \fig{fig:cir-multi-human}, we show the range profile squared magnitude for two human targets with a contiguous bandwidth of $4$~GHz (\fig{fig:cir-human-cont}) and with a non-contiguous bandwidth of the same aperture (\fig{fig:cir-human-gap}).
Even in this case, we observe the impact of grating lobes on the results of \ac{bp}, which is significantly mitigated by using \ac{spbp} instead.
Specifically, in \fig{fig:cir-human-gap} the two subjects represented by \ac{bp} as $4$ different peaks of different magnitude, while \ac{spbp} can correctly reconstruct two distinct peaks.

\begin{figure*}[t!]
    \centering
\includegraphics[width=\textwidth]{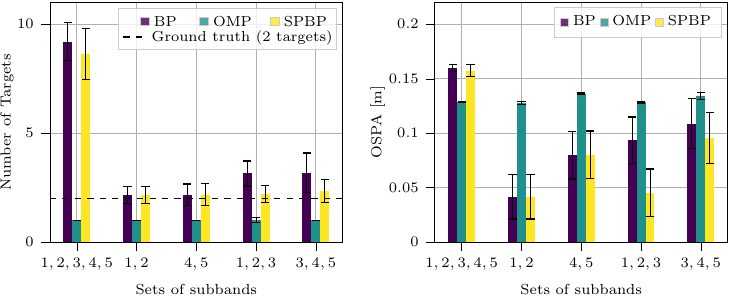} 
    \caption{Average number of targets (left) and \ac{ospa} (right) obtained by \ac{bp}, \ac{omp}, and the proposed \ac{spbp} on corner reflector and metal plate targets.
    The black whiskers represent one standard deviation. \ac{spbp} outperforms the other two algorithms by better reconstructing the correct number of targets and obtaining the lowest \ac{ospa}. When using the full set of subbands $S_1, \dots, S_5$, none of the algorithms can correctly detect the two objects.}
    \label{fig:ospa-corner}
\end{figure*}

\textbf{\ac{ospa} results:} 
In \fig{fig:ospa-corner}, we show the average number of targets and \ac{ospa} over $100$ tests obtained using two targets: a corner reflector and a metal plate.
Targets are located around $1.2$~m from the measurement device and spaced by $d_{\rm trg}= 0.2$~m. 
The ground truth target locations are obtained with a laser telemeter.
Since these measurements contain a systematic bias with respect to the exacted reflection points of the radio signal on the target, we align the final obtained range profiles to the ground truth with a rigid translation along the range dimension.

We use different combinations of the $5$ subbands $S_k$, with $k=1, \dots, 5$, and a cardinality penalty $\mu=d_{\rm trg}$, so that in the \ac{ospa} errors larger than the distance between the two targets are penalized as a missed detection.

\ac{omp} only detects one target in most of the tests. 
For this reason, it achieves the worst \ac{ospa} performance, as it is almost always penalized due to outputting the wrong cardinality for the set of targets.
The unsatisfactory performance of \ac{omp} is due to the fact that even a slight incoherence among the subbands introduces a model error in the \ac{omp} formulation, which assumes perfect coherence. 
This results in unpredictable effects in the \ac{omp} reconstruction and in a loss of resolution.

When using contiguous sets of subbands, i.e., $S_1$ and $S_2$ or $S_4$ and $S_5$, only $S_1$, \ac{spbp} is not used since the grating lobes due to non-contiguous subbands are not present.
Hence, in these cases, \ac{spbp} is represented with the same performance as \ac{bp}.

\ac{bp} works well with contiguous subbands, obtaining an accurate estimate of the number of targets and $5$ cm \ac{ospa} with $S_1 \cup S_2$, and $7.5$~cm \ac{ospa} with $S_4 \cup S_5$. 
This shows that the multiband range profile correctly reconstructs the number and locations of the two targets. 

Conversely, \ac{bp} has degraded performance when non-contiguous subbands are used.
Using $S_1 \cup S_2 \cup S_3$ and $S_3 \cup S_4 \cup S_5$ \ac{bp} detects on average $3.15\pm0.57$ and $3.18 \pm 0.90$ targets, respectively, showing that even a smaller frequency gap can lead to the wrong estimation of the number of targets.
The resulting \ac{ospa} reflects this, increasing with respect to the contiguous case to $7.15\pm 2.1$~cm and $9.33\pm 2.3$~cm, respectively.
\ac{spbp} instead estimates on average $2.20\pm 0.40$ and $2.34 \pm 0.53$ targets with $S_1 \cup S_2 \cup S_3$ and $S_3 \cup S_4 \cup S_5$, respectively, which well approximate the real number.
The resulting \ac{ospa} outperforms that of \ac{bp}

Using all subbands $S_1, \dots, S_5$ introduces strong grating lobes, caused by the non-contiguity of the subbands, degrading the performance of all methods.
In this case, \ac{bp} detects on average $9.19 \pm 0.87$ targets, with an \ac{ospa} of over $15$~cm.
\ac{spbp} only provides a slight gain but is also not able to estimate the correct number or targets, detecting $8.62\pm 1.17$ targets on average.

\begin{figure*}[t!]
	\begin{center}   
		\centering
        \includegraphics[width=\textwidth]{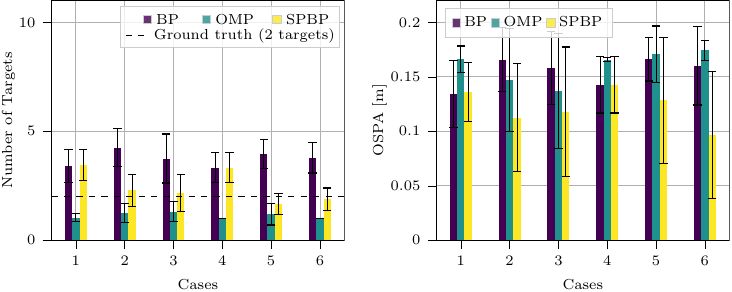} 
        \caption{Number of targets (left) and \ac{ospa} (right) obtained by \ac{bp}, \ac{omp}, and the proposed \ac{spbp} on two human targets. The black whiskers represent one standard deviation.
        \ac{spbp} outperforms the other two algorithms by better reconstructing the correct number of targets and obtaining the lowest \ac{ospa}.}
		\label{fig:ospa-humans}
	\end{center}
 \vspace{-0.5cm}
\end{figure*}

In \fig{fig:ospa-humans}, we show the mean number of targets and \ac{ospa} over $25$ tests obtained using two human targets.
The two subjects are located approximately $1$~m from the measurement device and spaced by $30$~cm.
Since human targets exhibit degraded coherence over very wide combined bandwidth, as shown in \secref{sec:coherence-quant}, we consider narrower frequency ranges as shown in the bottom plot in \fig{fig:allocations}.

Cases $1$ and $4$ represent contiguous $4$~GHz bandwidth allocations, which according to \fig{fig:human-coherence2} give a target coherence of above $0.75$.
From this, we expect humans to cause a single dominant peak in the multiband range profile, which would not be the case for a wider bandwidth combination.
The remaining cases are non-contiguous bandwidth allocations obtained by removing $1$~GHz bandwidth chunks, at different locations, from cases $1$ and $4$.

\fig{fig:ospa-humans} shows that \ac{bp} overestimates the number of targets in all cases, with an average of $3.75\pm0.31$ across all cases. 
Conversely, \ac{omp} has the same problem observed with the corner reflector and metal plate, i.e., it can not resolve the two targets in most cases.
Notably, since the subjects are more spaced apart than in the corner reflector and metal plate case, \ac{omp} performs slightly better. 
\ac{spbp} obtains a more accurate estimation of the real number of targets, with an average of $2.46\pm0.68$ detected targets over all cases.

\fig{fig:ospa-humans} confirms that \ac{spbp} outperforms both \ac{bp} and \ac{omp} on our measurements.
Averaging over all tests, \ac{spbp} has an \ac{ospa} of $12.25\pm1.54$~cm, compared to the $15.45\pm1.19$ and $16.01\pm1.34$~cm of \ac{bp} and \ac{omp}, respectively.
\fig{fig:ospa-humans} also agrees with the results in \fig{fig:human-coherence2}, showing that humans are less coherent at higher frequencies.
This can be seen from the fact that the \ac{ospa} in cases $4-6$ (frequencies from $14$ to $18$~GHz) of \ac{bp} is slightly higher than that of cases $1-3$ (frequencies from $7$ to $11$~GHz).

\section{Discussion}\label{sec:discussion}

In this work, we experimentally characterize the frequency dependency of the scattering response of \ac{isac} targets for coherent multiband sensing algorithms in \ac{fr3} ($6$-$22$~GHz).
We propose three new metrics to evaluate the coherence of the target response over the considered bandwidth: the \acf{mpc}, the \acf{nmpm}, and the \acf{empw}.
We test and validate these metrics on multiband scattering response data of a corner reflector, a metal plate, and human targets, demonstrating their effectiveness in capturing key aspects of coherent multiband ranging quality.
We assess the impact of using a fragmented spectrum and the consequent grating lobes on the resolution of multiple targets.
To mitigate the impact of grating lobes, we propose a heuristic algorithm (\ac{spbp}) and experimentally demonstrate its superiority to \ac{bp} and \ac{omp}, using \ac{fr3} bands considered by \ac{3gpp}, in terms of target detection and \ac{ospa}.

Our results demonstrate that the high fractional bandwidth of \ac{fr3} frequencies is a critical aspect to take into account when applying coherent multiband \ac{isac}.

The \textbf{main takeaways} are summarized as follows. 

    \noindent \textbf{1.} Different targets exhibit different \textbf{frequency-dependent behavior} and have coherent responses over different carrier frequencies and bandwidths. 
    
    \noindent \textbf{2.} While corner reflectors and flat metal surfaces are coherent over wide frequency ranges (almost the whole \ac{fr3}), \textbf{more complex targets} such as humans have an \textbf{incoherent response} if more than $3$-$4$~GHz bandwidth is used.
    
    \noindent \textbf{3.} Our experiments suggest that \textbf{higher carrier frequencies} (above $14$ GHz) lead to a \textbf{less coherent response} for human targets.
    The carrier frequency dependency of the target response aspect is not considered in the current \ac{isac} literature but is here shown to be significantly impacting the multiband combination result.
    Further research is needed on this aspect to determine whether this is a characteristic of the target or a consequence of slight movement of the person during the frequency sweep time, which leads to a larger phase variation at higher frequencies. 
    
    \noindent \textbf{4.} Using the currently considered allocations for wireless networks in \ac{fr3} by \ac{3gpp}, one must carefully consider the impact of grating lobes on coherent multiband ranging.
    Our results indicate that, \textbf{even for coherent targets} (e.g., corner reflectors or metal plates), combining all the considered non-contiguous subbands over a wide bandwidth leads to a \textbf{degraded range profile} due to the grating lobes.
    Therefore, it is key to design algorithms that mitigate the grating lobes to fully reap the benefit of the wide total frequency aperture in \ac{fr3}.
    
    \noindent \textbf{5.} Our results with the proposed \ac{spbp} algorithm suggest that (i)~using non-contiguous bands with \textbf{intermediate aperture sizes}, and (ii)~combining \textbf{different subsets} of the available subbands in a \textbf{non-linear} fashion to cancel out the grating lobes represent interesting research directions to enable coherent multiband ranging in \ac{fr3}.

Most of the existing radar literature on target anisotropy has focused on \textit{spatial anisotropy}, see, e.g.,~\cite{kim2003detection, yuan2024experimental, huang2021anisotropic}, using a single frequency band.
The frequency anisotropy of targets has been studied in~\cite{carriere1992high, huang2021anisotropic, patel2018review, bouwmeester2024effect, beasley2023multi}. 
These works have applied the Prony model~\cite{carriere1992high} or \ac{gtd}~\cite{keller2016geometrical, Moore2017UsingPhaseRadar} to explicitly model the frequency dependency of the complex-valued \ac{rcs} of the targets.
However, these models are only accurate for targets with predefined shapes, such as curved surfaces, edges, corners, flat plates, etc. 
This limits their usability in practice for complex targets such as humans and does not provide a direct insight on the coherence of targets over wide frequency bands.
In \ac{uwb} radar~\cite{Cheraghinia2024UWBOverview} instead, frequency anisotropy is typically not modeled, and the observed complex-valued \ac{rcs} of the targets over very wide contiguous bands is considered an average response over the total bandwidth.

In the \ac{isac} literature, fewer works have tackled this aspect by either experimentally measuring the real-valued \ac{rcs} of targets in multiple bands~\cite{azim2025statistical}, or using neural networks to learn the target response across a wide frequency aperture~\cite{li2025enabling} without modeling phase coherence explicitly.

The frequency anisotropy characterization done in this work takes a new and different approach compared to the above papers. 
Rather than directly modeling the behavior of the scattering phase, we design new practical metrics that can be used to \textit{assess} the coherence of a target's response in multiple bands either before (\ac{mpc}) or after applying multiband combination (\ac{nmpm}, \ac{empw}).
The proposed metrics provide insights into when it is feasible or convenient to use multiband processing with data collected from practical \ac{isac} devices, especially in high fractional bandwidth scenarios like \ac{fr3}.

Our result regarding the decreasing coherence of human targets with increasing carrier frequencies within the range $6$-$22$~GHz, given the same combined bandwidth, has not been observed before in the \ac{isac} literature, which has previously considered coherence to be independent of the carrier frequency~\cite{pegoraro2024hisac, li2025enabling}.
This encourages further research to characterize the coherence in the sub-6~GHz and \ac{mmwave} frequency ranges.

Several works in the literature address multiband coherent processing for radar \cite{cuomo1999ultrawide, Cuomo1992BandwidthExtrapolation, Borison1992SuperResolution, Stoica2009MissingDataIAA, gupta1994data, zhang2014coherent, wang2017multiband} or \ac{isac} \cite{liu2025carrier, liu2024integrated, pegoraro2024hisac, wei2024carrier,geng2025phase, li2025enabling} using non-contiguous bands.
These works assume frequency isotropic target response~\cite{pegoraro2024hisac, Stoica2009MissingDataIAA, geng2025phase} or use \ac{gtd}~\cite{cuomo1999ultrawide, zhang2014coherent} to simplify the frequency dependence of the complex-valued \ac{rcs}.
The employed algorithms to compensate for the non-contiguous bandwidth rely on all-pole modeling of the signal in the two bands~\cite{cuomo1999ultrawide, Cuomo1992BandwidthExtrapolation, wang2017multiband} or \ac{cs}~\cite{pegoraro2024hisac, li2025enabling, liu2024integrated}.
Both approaches suffer from high computational complexity and \textit{model mismatch} in case the target is not isotropic or does not satisfy the \ac{gtd} model, which significantly degrade the performance.
Our results specifically show that the \ac{omp} \ac{cs} algorithm is unable to resolve two targets in most cases, due to slight incoherence among different subbands and the consequent model mismatch.

In \cite{liu2025carrier}, the incoherence of targets is compensated for, eliminating phase differences in the scattering response due to frequency anisotropy. 
Although this allows multiband processing of the multiband data even under incoherence of the target, the resulting output may be misleading since it fails to capture intrinsic frequency-dependent features of the targets and reconstructs a single synthetic scattering point that may not represent a physical scattering point. 

The approach used in this paper is instead to limit the aperture of the considered multiple subbands to a frequency range where the target is coherent according to the proposed metrics. 
Then, we propose the \ac{spbp} algorithm with limited computational complexity that does not require explicit modeling of the target phase response as a function of frequency and mitigates the grating lobes. 

This work also fills a gap in terms of experimental measurements for coherent multiband \ac{isac} in \ac{fr3}.
Previous work has extensively provided incoherent radar \ac{rcs} measurements in the S, X, Ku, and K bands~\cite{scotti2015multi, balajti2006rcs, fioranelli2016experimental, bouwmeester2024effect}.
Notably, these works do not address the \textit{combination} of multiple bands in a coherent fashion, which is the key problem considered in this paper.
In the \ac{isac} literature, some works have characterized the communication channel in \ac{fr3}~\cite{Shakya2024FR1C_FR3_Indoor, serghiou2025indoor, azim2025statistical}, the real-valued \ac{rcs} of targets~\cite{serghiou2025indoor, fenollosa2024frequency, azim2025statistical, kodra2025multiband, zhang2025unified}, or the \ac{isac} channel \cite{bomfin2024experimental}.
However, these works do not analyze the \textit{phase coherence} of targets in \ac{fr3}, which is critical for coherent multiband \ac{isac}. Moreover, most of the above studies consider a limited set of carrier frequencies and bandwidths in \ac{fr3}, e.g., only $6.5$~GHz and $8.5$~GHz in~\cite{bomfin2024experimental}.
Other works~\cite{saberinia2008ranging, kazaz2022delay, noschese2021multi} have tackled multiband active localization where the objective is to compute the range of a communication device. 
Since no passive targets are involved, frequency anisotropy does not pertain to these works, and the proposed approaches do not model it.

To the best of our knowledge, this is the first study providing an evaluation of coherent multiband processing over almost the \textit{whole} \ac{fr3} ($6$-$22$~GHz), with a focus on coherent multiband \ac{isac} algorithms.
The obtained results shed light on key aspects of multiband \ac{fr3} \ac{isac} and represent a first step towards designing multiband algorithms for \ac{6g} wireless networks and planning spectrum allocation for maximum coherent multiband aggregation performance.

Further research should incorporate the concept of frequency anisotropy into the design of algorithms for coherent multiband aggregation.
For example, the experimental characterization of the scattering phase from complex targets could be leveraged to build statistical models, used as inputs for new aggregation algorithms.
Data-driven approaches stand out as promising candidates to extract non-linear frequency-dependent target scattering properties in this sense.

Moreover, further investigation of the \ac{spbp} approach for grating lobes mitigation should be conducted, addressing the attenuation of weaker targets with respect to dominating ones. 
This aspect represents a challenging problem in complex \ac{isac} scenarios with multiple targets having \acp{rcs} with orders of magnitude differences.  

Finally, the target anisotropy characterization should be extended to the \textit{spatial dimension}, evaluating the phase coherence for different incidence angles of the \ac{isac} signals.
Indeed, complex extended targets may exhibit different scattering properties at different angles, which is not addressed in the \ac{isac} literature~\cite{tagliaferri2024cooperative}.
This aspect is critical to enable coherent \ac{isac} processing over multiple distributed nodes, such as different base stations or access points.

\section{Methods}\label{sec:methods}

In this section, we describe the methodology used in this work, including the mathematical system model, the description of the algorithms used, and details on the experimental data collection.

\subsection{System and signal models}\label{sec:sysmodel}

In this work, we consider a multiband \ac{siso} system with a transmitter (Tx) antenna and a receiver (Rx) antenna, which share the same \ac{lo}. The \ac{siso} system operates on $K$ subbands, each centered on a carrier frequency $f_k$ and of bandwidth $B_k$. We collect the carrier frequencies and bandwidths in sets $\mathcal{F} = \{f_k\}_{k=0}^{K-1}$, and $\mathcal{B} = \{B_k\}_{k=0}^{K-1}$, respectively. 
We assume that $L$ \textit{static} scattering centers are present in the environment, indexed by $\ell=1,\dots,L$, located at distances $R_\ell$ from the \ac{siso} ISAC terminal. 
Each sensing target may be composed of a single scattering center or \textit{multiple} scattering centers (extended target, for example, a human body), which are only resolved by the \ac{isac} system if sufficient bandwidth is available.
Therefore, the number of sensing targets is, in general, lower than or equal to $L$.

Each scattering center is characterized by a complex \textit{frequency-dependent} \ac{rcs}, $\rho_{\ell,k}e^{j\theta_{\ell,k}}$, whose magnitude $\rho_{\ell,k}$ and phase $\theta_{\ell,k}$ contain the effect of the combined path loss and scattering attenuation, and of the scattering phase response, respectively.
Obtaining an explicit model for $\rho_{\ell,k}e^{j\theta_{\ell,k}}$ as a function of frequency is hard. 
The phase of the $\ell$-th scattering center, in general, is a function of its physical composition, namely material, wave polarization, angles of incidence and observation, and frequency of operation. We characterize scattering centers from a single monostatic angle of observation and a single polarization, observing the implicit relation between the target's material, shape, and carrier frequency.
Existing works have found models that hold for specific classes of ideal scatterers using \ac{gtd}~\cite{keller2016geometrical}.
In this paper, instead, we do not explicitly model $\rho_{\ell,k}e^{j\theta_{\ell,k}}$ and rather evaluate its variation across frequency indirectly, using the metrics in \secref{sec:metrics}.

We stress that while existing multiband \ac{isac} works model the \ac{rcs} as constant across frequency in both magnitude and phase~\cite{pegoraro2024hisac, li2025enabling}, this is the first \ac{isac} work to consider the frequency anisotropy of the targets and its impact on multiband processing.

The \ac{siso} \ac{isac} system emits an \ac{ofdm} signal over the $K$ subbands, each subband having $N_k$ subcarriers and fixed subcarrier spacing $\Delta_f$. 
The propagation delay from the $\ell$-th scattering center is $\tau_\ell = 2 R_\ell /c $, with $c$ being the speed of light and $R_\ell$ the distance between the scattering center and the \ac{isac} device, respectively. 
To perform a frequency sweep of multiple subbands, one \ac{ofdm} symbol per subband is emitted sequentially in time, as detailed in \secref{sec:freq-sweep}. 

The baseband signal for the $k$-th \ac{ofdm} symbol (hence for the $k$-th subband), assumed to have unit-power, is
\begin{equation}\label{eq:Tx_signal_time_bb}
    x_{k}(t) = \sum_{n = -\frac{N_k}{2}}^{\frac{N_k}{2}-1} a_{n} e^{j 2 \pi  n \Delta_f t}, \quad \quad \mbox{for  }  t\in \left[kT , kT + T\right]
\end{equation}
where $a_{n}\in \mathbb{C}$ is the frequency domain sequence of known preamble pilots, equal for all subbands, $N_k$ is the number of subcarriers for subband $k$, and $T=1/\Delta_f$ is the duration of the \ac{ofdm} symbol. 
The pass-band Tx signal is 
\begin{equation}\label{eq:Tx_signal_time_pb}
    x^{\rm RF}_{k}(t) = x_k(t) e^{j2\pi f_k t}.
\end{equation}

The Tx signal propagates through the environment and undergoes scattering from the $L$ scattering centers. 
The \ac{siso} \ac{cir} is defined as follows  
\begin{equation}
    h_{k}(t) = \sum_{\ell=1}^L \rho_{\ell,k} e^{j \theta_{\ell,k}} \delta(t-\tau_\ell),
\end{equation}
where $\delta(\cdot)$ is the Dirac delta. 
The scattering amplitude is a function of the distance and of the \ac{rcs}, $\sigma_{\ell, k}$, given by the \textit{radar equation}~\cite{richards2010principles}
\begin{equation}
   \rho_{\ell, k} = \sqrt{\frac{c^2 G_{k,\rm tx} G_{k,\rm rx} \sigma_{\ell, k}}{f_k^2(4 \pi)^3 R_\ell^4}}
\end{equation}
where $G_{k,{\rm tx}}, G_{k,{\rm rx}}$ are the frequency-dependent antenna gains at the Tx and Rx, respectively. 

The Rx signal in the time domain on the $k$-th subband (thus the $k$-th \ac{ofdm} symbol) is the convolution of the Tx signal and the \ac{cir}.
The Rx then introduces timing errors due to the non-ideality of the \ac{lo}.
We define the timing error on subband $k$ at time $t$ as~\cite{Loria2023OscillatorPhaseNoiseSAR}
\begin{equation}
    \epsilon_k(t) = \epsilon_{k,0} + \epsilon_{k,1} t + \epsilon_{k,\rm rnd}(t)
\end{equation}
where $\epsilon_{k,0}$ is the \ac{to}, $\epsilon_{k,1}$ is the normalized \ac{cfo}, and $\epsilon_{k,\rm rnd}(t)$ is the random timing error due to the phase noise of the \ac{lo}.

Defining $ \Delta\epsilon_{k,\rm rnd}(t, \tau_{\ell}) = \epsilon_{k,\rm rnd}(t-\tau_\ell) - \epsilon_{k,\rm rnd}(t)$, the passband Rx signal affected by timing errors can be written as
\begin{equation}\label{eq:Rx_signal_time}
\small
\begin{split}
   y^{\rm RF}_{k}(t) 
   & = \sum_{\ell=1}^L \rho_{\ell,k} e^{j \theta_{\ell,k}}  x^{\rm RF}_{k}(t - \tau_\ell + \epsilon_k(t-\tau_\ell) - \epsilon_k(t))+ z_k(t)\\
   & = \sum_{\ell=1}^L \rho_{\ell,k} e^{j \theta_{\ell,k}}  x^{\rm RF}_{k}(t - \tau_\ell + \epsilon_{k,1}\tau_\ell + \Delta\epsilon_{k,\rm rnd}(t, \tau_{\ell}))\\
   &+ z_k(t)\\
   &= \sum_{\ell=1}^L \rho_{\ell,k} e^{j \theta_{\ell,k}} \sum_{n = -\frac{N_k}{2}}^{\frac{N_k}{2}-1} a_{n} e^{j 2 \pi ( f_k + n \Delta_f)t} e^{-j2\pi ( f_k + n \Delta_f) \tau_\ell}   \\
   &  \quad \cdot e^{-j2\pi ( f_k + n \Delta_f)\epsilon_{k,1}\tau_\ell}
   e^{j2\pi ( f_k + n \Delta_f) \Delta\epsilon_{k,\rm rnd}(t, \tau_{\ell})} + z_k(t) .    
\end{split}
\end{equation}
The signal is corrupted by non-linear clock errors and by an additive white Gaussian noise term, $z_k(t)\sim \mathcal{CN}(0,\sigma_z^2)$. 
Notably, since we consider a monostatic system, the \acp{to} cancel out in \eq{eq:Rx_signal_time}.
To simplify the above expression, we make the following assumptions: \textit{(i)} the \ac{cfo} does not significantly affect the Rx signal, as it is typically $\epsilon_{k,1} \ll 1$ (for RF clocks, $\epsilon_{k,1}\sim \mathcal{N}(0,10^{-6}$) and, for distances in the order of meters to tens of meters and \ac{fr3} frequencies, the resulting carrier phase rotation $e^{-j2\pi f_k \epsilon_{k,1}\tau_\ell}$ is negligible \cite{tagliaferri2024cooperative}; 
\textit{(ii)} the random time error is much less than the typical pulse duration, i.e., $|\epsilon_{k,\rm rnd}(t-\tau_\ell) - \epsilon_{k,\rm rnd}(t)| \ll 1/B_k$, always verified in practice;
\textit{(iii)} the phase rotation due to differential phase noise $\varphi_k(\tau_\ell) = 2 \pi f_k \Delta\epsilon_{k,\rm rnd}(t, \tau_{\ell})$ is negligible in the considered monostatic setup. 
This latter assumption is motivated by applying standard modeling of the \ac{lo} as a random process with a power law spectrum to our scenario. 
For typical observation times of tens/hundreds of microseconds and typical ranges of tens of meters (propagation delays $\tau_\ell$ in the order of tens of nanoseconds), the differential phase noise $\varphi_k(\tau_\ell)$ is only due to the white component of the phase spectrum of the LO $S_\phi(\nu)$, namely its variance is 
 $\sigma^2_{\varphi_k}(\tau_\ell) \simeq \left(f_k / \overline{\nu}\right)^2 2 \alpha_0 B_k$~\cite{TI_PLL_Basics} 
where $\alpha_0$ is the power spectral density of the white component of the phase noise and $\overline{\nu}$ is the nominal frequency of the LO.
For good clocks, $\alpha_0$ is low enough to neglect the impact of phase noise. For the considered HW setup, $\alpha_0 = -210$ dBc/Hz, $B_k=1$ GHz, $f_k = 22$ GHz (max) and $\overline{\nu}=10$ MHz, thus $\sigma_{\varphi_k}(\tau_\ell) \approx 0.2$ deg. This latter derivation justifies neglecting the phase noise in the work hereafter. In any case, the power of the phase noise after matched filtering (CIR estimation) is further reduced by the time-bandwidth product $T B_k = N_k$ (number of subcarriers). For low-end clocks, characterized by a higher intrinsic phase noise, the only opportunity is to employ multiple \ac{ofdm} symbols per subband and average the estimated \acp{cir}.

Using the above simplifications, the baseband Rx signal after downconversion is 
\begin{equation}\label{eq:Rx_signal_time_down}
\begin{split}
   y_{k}(t) 
   = y^{\rm RF}_{k}(t) e^{- j 2 \pi f_k t} =& \sum_{\ell=1}^L \rho_{\ell,k} e^{j \theta_{\ell,k}} x_k(t - \tau_\ell) e^{-j2\pi f_k \tau_\ell} \\
   & + z_k(t) 
\end{split}
\end{equation}
In the next section, we detail the preprocessing steps applied to the baseband Rx signal to prepare it for coherent multiband combination.

\subsection{Pre-processing of the Rx signal}

The RF signal in \eq{eq:Rx_signal_time_down} is pre-processed before the extraction of the target's coherence metrics. The goal is to obtain the \ac{cir} for each subband, which requires the following steps:
\begin{enumerate}
    \item Sampling with sampling interval $1/B_k$ and conversion to the frequency domain using a \ac{dft}, obtaining
    \begin{equation}
    \begin{split}
        Y_k(n \Delta_f) & = X_k(n \Delta_f) \sum_{\ell=0}^L \rho_{\ell,k} e^{j \theta_{\ell,k}} e^{- j 2 \pi \left(f_k + n\Delta_f \right)\tau_\ell} \\
        & + Z_{k}(n \Delta_f)
    \end{split}
    \end{equation}
    where $X_k(n \Delta_f)$ is the frequency domain Tx signal and $Z_k(n \Delta_f)$ is the noise in the frequency domain. 
    Note that subsequent multiband processing will require higher time granularity than $1/B_k$, thus requiring interpolation. 
\item Estimation of the \ac{siso} channel. 
The ideal \ac{cfr} in subband $k$ is
\begin{equation}\label{eq:cfr-real}
\begin{split}
         H_{k}(n \Delta_f) = \sum_{\ell=0}^L \rho_{\ell,k} e^{j \theta_{\ell,k}} e^{- j 2 \pi \left(f_k + n\Delta_f \right)\tau_\ell}.
\end{split}
\end{equation}
In practice, we estimate the \ac{cfr} in the frequency domain through elementwise division of the Rx symbols and the known Tx ones as
\begin{equation}\label{eq:cfr-est}
    \widetilde{H}_{k}(n \Delta_f) = \frac{Y_k(n \Delta_f)}{ X_k(n \Delta_f)} = H_{k}(n \Delta_f) H_{k}^{\rm hw}(n\Delta_f) + W_{k}(f),
\end{equation}
where $H_{k}^{\rm hw}(n\Delta_f)$ is the frequency response of the measurement device and $W_{k}(f) = Z_{k}(f)/X_k(n \Delta_f)$.

\item Hardware calibration. This step is needed to compensate for the frequency response of the measurement device in subband $k$, $H_{k}^{\rm hw}(n\Delta_f)$, which may introduce distortion in the phase of the estimated channel.
We estimate $H_{k}^{\rm hw}(n\Delta_f)$ using the procedure detailed in the supplementary material (Supplementary note - 3).
The calibration step obtains a calibrated version of the \ac{cfr} in \eq{eq:cfr-est} as
\begin{equation}\label{eq:cfr-cal}
\begin{split}
         \widehat{H}_{k}(n \Delta_f) = \widetilde{H}_{k}(n \Delta_f) / H_{k}^{\rm hw}(n\Delta_f),
\end{split}
\end{equation}
which removes undesired phase variations due to the non-ideal hardware response across frequency.

\item Estimation of the SISO CIR by computing the \ac{idft} of $\widehat{H}_k(n \Delta_f)$
\begin{equation}\label{eq:cir-est}
\begin{split}
     \widehat{h}_{k}(t) = \sum_{\ell=0}^L \rho_{\ell,k} e^{j \theta_{\ell,k}} \mathrm{sinc}\left[B_k(t-\tau_\ell)\right]  e^{- j 2 \pi f_k \tau_\ell} + w_k(t),
     \end{split}
\end{equation}
where $w_k(t)$ is the \ac{idft} of the noise, and $\mathrm{sinc}(x) = \sin(\pi x) / (\pi x)$ is the cardinal sine function.
In \eq{eq:cir-est} we use the continuous time variable $t$ for the \ac{cir} although in practice this is discretized in the output of the \ac{idft}.
Our choice of using $t$ stems from the fact that we compute the \ac{idft} on a much denser time grid than the original sampling interval $1/B_k$, approximating the continuous-time \ac{cir}.
This is done because the \ac{bp} algorithm, detailed in the next section, requires evaluating the \ac{cir} at arbitrary delay values that may not lie on the coarse grid determined by the original sampling interval.

\eq{eq:cir-est} is a collection of scaled and shifted cardinal sine functions affected by the scattering phase $\theta_{\ell,k}$ of the $\ell$-th scattering center in subband $k$.
Moreover, an additional phase term due to the propagation of the carrier is present, $-2 \pi f_k \tau_\ell$, which prevents a direct combination of the \acp{cir} since it introduces a subband-dependent phase shift for that same delay $\tau_\ell$.
The resolution of \eq{eq:cir-est} in distinguishing multiple scattering centers is limited by the bandwidth of a single subband.
Specifically, the resolution is given by the width of the main lobe of the sinc functions, which is $1/B_k$ in delay units and $c/(2B_k)$ in range units.

\end{enumerate} 

\subsection{Multiband combination algorithms}\label{sec:mb-algorithms}

In this section, we detail the multiband combination algorithms used in this work, namely \ac{bp} and the proposed \ac{spbp}.
\ac{omp} is described in detail in the supplementary material (Supplementary note - 2).

\subsubsection{Backprojection algorithm and multiband \ac{raf}}\label{sec:bp-raf}

To increase the resolution of the estimated \ac{cir} using $K$ subbands we adopt the time-domain \ac{bp} algorithm used in~\cite{tagliaferri2024cooperative}, adapting it to one-dimensional multiband ranging. 
The pre-processing provides the \acp{cir} of the subbands, which are still affected by the propagation carrier phase. 
The first step in \ac{bp} is to isolate the target's scattering phase by compensating for the propagation-dependent carrier phase term $2\pi f_k\tau_\ell$.
This is done by expressing the propagation delay $\tau_\ell$ as a function of the scattering center's distance, i.e., $\tau_\ell = 2R_\ell /c$, and computing
\begin{equation}\label{eq:prop-phase-comp}
\begin{split}
   \eta_{k}(R) & =  \widehat{h}_{k}\left(\frac{2 R}{c}\right)e^{j \frac{4 \pi f_k}{c} R } \\
   & = \sum_{\ell=0}^L \rho_{\ell,k} e^{j \theta_{\ell,k}} \underbrace{\mathrm{sinc}\left[\frac{2B_k}{c} (R-R_\ell)\right]}_{\chi_k(R-R_\ell)} e^{j \frac{4 \pi f_k}{c}(R - R_\ell)} \\
   & + w_k(R)
\end{split}
\end{equation}
where $ w_k(R)$ is a noise term, and $\chi_k(R)$ is called \ac{raf} of the single subband and coincides with a sinc function that depends uniquely on bandwidth $B_k$. 
At $R = R_{\ell}$, the term $e^{j \frac{4 \pi f_k}{c} R }$ cancels out the carrier phase term, so that the phase of the $\ell$-th element in the sum equals the phase of the scattering center's phase response $\theta_{k, \ell}$.

As a second step, \ac{bp} sums the $K$ range profiles obtained by \eq{eq:prop-phase-comp} coherently as
\begin{align}\label{eq:bp-output-general}
    \eta(R) = & \sum_{k =0}^{K-1} \eta_{k}(R) \\= &\sum_{\ell=0}^L \sum_{k=0}^{K-1}  \rho_{\ell,k} e^{j \theta_{\ell,k}} \chi_k(R-R_\ell)e^{j \frac{4 \pi f_k}{c} (R - R_\ell)} + w(R),
\end{align}
where $ w(R)$ is a noise term.
\eq{eq:bp-output-general} represents the most general output of the \ac{bp} algorithm, when the complex \ac{rcs} of the target changes with frequency, and the carrier frequencies and bandwidths of the subbands are unequal.
In this case, the scattering responses of the single scattering centers can not be taken out of the sum over $k$, so the resulting multiband response is \textit{different for each target}.
From \eq{eq:bp-output-general}, it is not immediately evident why \ac{bp} achieves a resolution improvement with respect to single subband processing. 
Therefore, in the following, we simplify \eq{eq:bp-output-general} by making some assumptions to highlight the factors that contribute to the resolution improvement.

If the response of the scattering centers is constant across the considered frequency range (\textit{frequency isotropic}), i.e., $\rho_{\ell,k} e^{j \theta_{\ell,k}} = \rho_{\ell} e^{j \theta_{\ell}}$, $\forall k$, \eq{eq:bp-output-general} can be simplified. 
In this case, \eq{eq:bp-output-general} becomes
\begin{equation}\label{eq:bp-output-iso}
\begin{split}
    \eta_{\rm iso}(R) &=  \sum_{\ell=0}^L  \rho_{\ell} e^{j \theta_{\ell}}  \sum_{k=0}^{K-1} \chi_k(R-R_\ell)e^{j \frac{4 \pi f_k}{c} (R - R_\ell)} + w(R) \\
    & =  \sum_{\ell=0}^L \rho_{\ell} e^{j \theta_{\ell}} \Psi_{\mathcal{F}, \mathcal{B}}(R - R_\ell) + w(R),
\end{split}
\end{equation}
where $\Psi_{\mathcal{F}, \mathcal{B}}(R)$ is the ideal \ac{raf} for frequency isotropic targets and depends on the location and width of the subbands in the frequency domain.
Importantly, only assuming isotropy in the frequency domain one can write a single \ac{raf} that is \textit{independent} of the specific target.
\eq{eq:bp-output-iso} can be further simplified if the subbands are chosen with \textit{(i)} equal bandwidth, $B_k = B, \forall k$, or \textit{(ii)} equal bandwidth and equally spaced carrier frequencies, $B_k = B, \forall k$, $f_k = f_0 + k\Delta f_{\rm b}$ where $\Delta f_{\rm b}$ is the inter-carrier spacing.

In the first case (equal bandwidth) the single-band $\mathrm{sinc}$ function has equal width for all subbands and is denoted by $\chi(R)$. 
The range profile becomes
\begin{equation}\label{eq:bp-output-eqband}
\begin{split}
    \eta_{\rm EB}(R) &=  \sum_{\ell=0}^L  \rho_{\ell} e^{j \theta_{\ell}}  \chi(R-R_\ell)\sum_{k=0}^{K-1} e^{j \frac{4 \pi f_k}{c} (R - R_\ell)} + w(R) \\
    & =  \sum_{\ell=0}^L \rho_{\ell} e^{j \theta_{\ell}} \Psi_{\mathcal{F}}(R - R_\ell) + w(R),
\end{split}
\end{equation}
where the \ac{raf} now only depends on the single bandwidth $B$ and on the carrier frequencies $\mathcal{F}$.

In the second case (equal bandwidth and equally spaced carrier frequencies) the range profile is
\begin{equation}\label{eq:bp-output-eqspace}
\begin{split}
    \eta_{\rm ES}(R) &=  \sum_{\ell=0}^L  \rho_{\ell} e^{j \theta_{\ell}}  \chi(R-R_\ell)\sum_{k=0}^{K-1} e^{j \frac{4 \pi f_k}{c} (R - R_\ell)} + w(R) \\
    & =  \sum_{\ell=0}^L  \rho_{\ell} e^{j \theta_{\ell}}  \chi(R-R_\ell)\Lambda (R - R_\ell) + w(R) \\
    & =  \sum_{\ell=0}^L \rho_{\ell} e^{j \theta_{\ell}} \Psi(R - R_\ell) + w(R),
\end{split}
\end{equation}
where
\begin{equation}
\begin{split}
        \Lambda(R) = \sum_{k=0}^{K-1} e^{j \frac{4 \pi f_k}{c} R} 
        = e^{j \frac{\pi}{c} \left[2 f_{0} + (K-1)\Delta f_{\rm b}\right]R} \; \frac{\sin \left( \frac{2 \pi}{c} K \Delta f_{\rm b} R \right)}{\sin \left(\frac{2 \pi}{c} \Delta f_{\rm b} R)\right)}
\end{split}
\end{equation}
is a Dirichlet kernel whose main lobe width defines the theoretical multiband range resolution
\begin{equation}
    \rho_{R, \rm MB} = \frac{c}{2 K \Delta f_{\rm b}} ,
\end{equation}
related to the total bandwidth aperture of the multiband system, $B_{\rm tot} = K \Delta f_{\rm b}$.
This explains why \ac{bp} achieves a resolution improvement with respect to single subband processing by summing the \acp{cir} of the single subbands after removing the carrier phase contribution.
$\Lambda(R)$ has a peak in \mbox{$R=0$} and is periodic of period $c/(2 \Delta f_{\rm b})$. 
Therefore, by increasing the spacing among carrier frequencies above $ B/2$ (non-contiguous subbands), more than one repetition of the peak $\Lambda(0)$ falls inside the mainlobe of the single-band $\mathrm{sinc}$, $\chi(R)$. 
This mathematically explains the appearance of grating lobes for non-contiguous subbands.  

The consequence of the above properties is that the multiband system potentially has a resolution equal to a \textit{fullband} system with a contiguous bandwidth spanning from the smallest to the largest frequency in the spectrum of the Tx signals in all subbands.
However, the grating lobes degrade such resolution by introducing artifacts. 

\subsection{Subsets Product \ac{bp} algorithm}\label{sec:subsets-prod-bp}

We design a heuristic modification of the \ac{bp} algorithm to mitigate the impact of grating lobes in the \ac{raf}, called \ac{spbp}.
The key idea is that different sets of subbands will produce grating lobes at different locations, since the portions of the spectrum occupied by the subbands are different.
Conversely, targets will appear at the same ranges regardless of the subbands' location in the frequency domain.
We exploit this property to cancel out grating lobes by taking the \textit{product} of the magnitude of range profiles obtained using different sets of subbands.

Consider two different subsets of the set of indices of the available subbands $\mathcal{K}= \{1,\dots,K-1\}$, termed $\mathcal{K}_0$ and $\mathcal{K}_1$, respectively, and denote by $\Psi_{\mathcal{X}}(R)$ the multiband \ac{raf} obtained by combining the subbands whose indices are in set $\mathcal{X}$.
We select $\mathcal{K}_0$ in such a way that the corresponding subbands have the same total bandwidth as the original set of subbands, to ensure the nominal resolution is the same. 
This is done by including in $\mathcal{K}_0$ the first and last indices of the subbands, namely $0$ and $K-1$.
The remaining elements of $\mathcal{K}_0$ are selected randomly from $\mathcal{K}$, keeping a cardinality equal to $|\mathcal{K}_0| = K-1$.
This is done to include the maximum number of possible subbands, thus reducing the gaps in the frequency domain, without taking $\mathcal{K}_0 = \mathcal{K}$.

Once $\mathcal{K}_0$ is selected, we numerically compute the corresponding \ac{raf}
$\Psi_{\mathcal{K}_0}(R)$ as
\begin{equation}
     \Psi_{\mathcal{K}_0}(R) = \sum_{k\in \mathcal{K}}  \chi_k(R) e^{j \frac{4 \pi f_k}{c} R}.
\end{equation}
Then we construct all the possible subsets of $\mathcal{K}$, which are collected in the product set $\Pi(\mathcal{K})$.
\ac{spbp} searches for the set $\mathcal{K}_1 \in \Pi(\mathcal{K}) \setminus \mathcal{K} \setminus \mathcal{K}_0$ that has a \ac{raf} whose product with $|\Psi_{\mathcal{K}_0}(R)|$ (in magnitude) leads to the minimum \ac{pslr}~\cite{tagliaferri_wavefield}.
The \ac{pslr} evaluates the ratio between the peak of the \ac{raf} and the maximum magnitude of its sidelobes, intended as the set of peaks outside of the main lobe.

The \ac{raf} product of the two subsets is
\begin{equation}
    \Gamma_{\mathcal{K}_0, \mathcal{K}_1}(R) = |\Psi_{\mathcal{K}_0}(R)|\cdot |\Psi_{\mathcal{K}_1}(R)|,
\end{equation}
and the minimization of the \ac{pslr} can then be written as
\begin{equation}\label{eq:k-min}
    \mathcal{K}_1 = \min_{\mathcal{K}' \in \Pi(\mathcal{K}) \setminus \mathcal{K} \setminus \mathcal{K}_0} \left(\frac{|\Gamma_{\mathcal{K}_0, \mathcal{K}'}(0)|}{\max_{R \in [-R_{\max}, R_{\max}] \setminus \Omega} |\Gamma_{\mathcal{K}_0, \mathcal{K}'}(R)|}\right),
\end{equation}
where $R_{\max}$ is the maximum range of interest and $\Omega$ is the main lobe region of the \ac{raf}. 
\eq{eq:k-min} is solved by exhaustive search since $|\mathcal{K}|$ is small in practice due to the non-ideal coherence of the target, which limits the number of subbands that can be combined.
Further conditions can be imposed on the candidate sets considered in the search to speed it up, e.g., restricting to subsets with at least a certain number of elements or that have at least a certain spectral coverage. 

The final magnitude of the range profile obtained with \ac{spbp} is
\begin{equation}\label{eq:spbp-rp}
    |\eta_{\rm SPBP}(R)| = \left|\sum_{k \in \mathcal{K}_0}\eta_{k}(R)\right| \cdot \left|\sum_{k \in \mathcal{K}_1} \eta_{k}(R)\right|.
\end{equation}
One drawback of \ac{spbp} is that the product in \eq{eq:spbp-rp}, besides strongly attenuating sidelobes, also relatively attenuates weak scatterers with respect to stronger ones.
Therefore, if some targets are significantly weaker than others, they may be canceled out and become undetectable.
In our results, this phenomenon was not significant, as shown in \secref{sec:resolution-results}, but further research is needed to mitigate this problem.

\subsection{Measurement system and experimental protocol}\label{sec:setup}

In this section, we detail our experimental setup for multiband ranging, including details on the measurement system, targets, hardware calibration procedures, and measurement protocol.

\subsubsection{FR3 measurement system}\label{sec:fr3-sys}

To implement our experiment in FR3, we utilize a \ac{sdr} platform in monostatic configuration, using 1 transmit and 1 receive chains. 
The \ac{sdr} includes the following components

\textbf{Digital baseband RFSoC:} A Xilinx \ac{rfsoc} 4x2 Kit equipped with a ZU48DR processor is used for baseband signal processing, including digital-to-analog (and vice-versa) conversion. The board is capable of generating \ac{rf} signals up to $6$~GHz, and therefore cannot directly generate \ac{fr3} signals. It is instead utilized to generate an \ac{if} signal centered at $1$~GHz that is up/down converted in a later stage.

\textbf{\ac{rf} transceiver Pi-Radio board:} The \ac{if} signal generated by the \ac{rfsoc} \mbox{$4\times 2$} Kit is up/down converted by a Pi-Radio TRX board, which translates the fixed-frequency \ac{if} to a configurable frequency in the $6$-$24$~GHz range, enabling the frequency sweep. The up and down conversions are performed coherently, ensuring consistent phase measurements. More details
of the board can be found in \cite{mezzavilla2024frequency}.

\textbf{Vivaldi wideband antenna:} The measurements are performed over a bandwidth of over $15$~GHz centered around $14$~GHz, which constitutes a fractional bandwidth greater than $100$\%. Such a fractional bandwidth is difficult to achieve with standard antennas such as patches or dipoles, which usually provide a percentage fractional bandwidth in the order of units to a few tens. To address this, we use a wideband Vivaldi antenna which, thanks to its exponentially tapered structure, can operate across the whole $6$-$24$~GHz band.

\begin{figure*}[t!]
    \centering
    \includegraphics[width=0.8\textwidth]{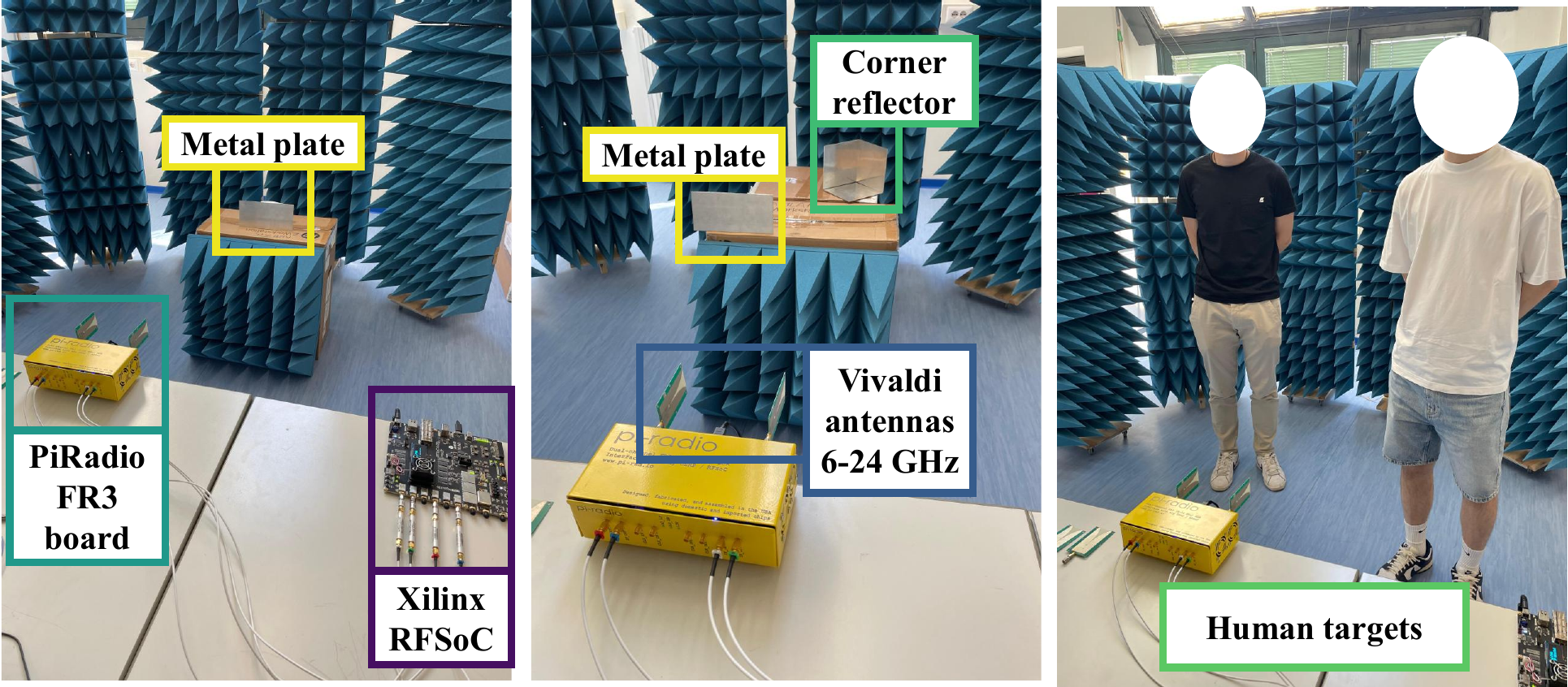}
    \caption{Experimental testbed and considered targets. The Xilinx \ac{rfsoc} generates the baseband signal and sends it to the Pi-Radio board for \ac{fr3} upconversion. The Vivaldi antennas enable wideband signal transmission.
    The different types of considered targets, namely the corner reflector, metal plate, and humans are further described in \secref{sec:targets}.}
    \label{fig:testbed}
\end{figure*}

To validate our experiments and obtain reference data to be used as a sanity check, we also adopt a \ac{vna} configured to operate in the $6$-$24$ GHz and connected to the same Vivaldi antenna used on the \ac{sdr} platform.
The device is calibrated, so it does not introduce phase distortion due to non-idealities of the hardware.
The \ac{vna} is a N9952B FieldFox Handheld Microwave Analyzer from Keysight, and can operate coherently over $50$~GHz of bandwidth.
We collect our measurements in a laboratory environment where we cover walls, furniture, and the support of the targets with panels made of \ac{rf} absorbing material, as shown in \fig{fig:testbed}. 

\subsubsection{Considered targets}\label{sec:targets}

In our experimental results, we consider the following three types of targets.
\begin{itemize}
    \item \textbf{Corner reflector:} We use a corner-cube reflector with $15$~cm side length, as shown in the middle picture in \fig{fig:testbed}. The corner reflector is calibrated for radar applications.
    \item \textbf{Metal plate:} We use a flat metal plate of dimensions $25 \times 10$~cm (\fig{fig:testbed}). The thickness of the plate is a few millimeters.
    \item \textbf{Static human:} Human subjects are instructed to stand as still as possible in front of the measurement device. 
    To mitigate the impact of respiration on the phase measurements, they are asked to hold their breath for $2$-$3$ seconds during the data collection, which lasts $150$~ms for the full frequency sweep.
\end{itemize}

\subsubsection{Frequency sweep}\label{sec:freq-sweep}

To obtain coherent \ac{cfr} estimates over the bandwidth $6$-$22$ GHz, we use the Pi-Radio FR3 board to implement a frequency sweep across the full bandwidth. 
This is done by transmitting random \ac{ofdm} pilot signals from a $4$-\ac{qam} over a configurable bandwidth centered around a configurable carrier frequency
The carrier frequency is changed in subsequent \ac{ofdm} symbols. The switching between different carrier frequencies requires $t_{\rm switch}= 10$ ms in the Pi-Radio board, so this is the minimum timing between \ac{ofdm} symbols that can be configured in our measurements.

In our experiments, we use two different bandwidth values $0.5$ or $1$~GHz, depending on the specific experiment.
We select the carrier frequencies to obtain a \textit{contiguous} set of \ac{cfr} estimates. When using $0.5$~GHz, we select the set $\{6.5, 7, \dots, 22\}$~GHz, with $32$ subbands, while using $1$~GHz we select $\{6.5, 7.5, \dots, 21.5\}$~GHz, with $15$ subbands. 
The different cardinality of the two sets has the important consequence that the total duration of the frequency sweep using $0.5$~GHz is $32t_{\rm switch} = 320$ ms, while using $1$~GHz it is $15t_{\rm switch} = 150$~ms. For this reason, we use $1$~GHz in the measurements involving human targets since a shorter total measurement time is preferable to avoid incoherence due to small involuntary movements of the person.

\bibliography{biblio}
\bibliographystyle{IEEEtran}
\end{document}